\documentclass[amsmath,amssymb,superscriptaddress, showpacs, showkeys,twocolumn,prx]{revtex4-1}

\usepackage{graphicx}
\usepackage{dcolumn}
\usepackage{bm}
\usepackage{amsmath}
\usepackage{amssymb,amsthm}
\usepackage{color}
\usepackage{epstopdf} 
\usepackage[sans]{dsfont}
\usepackage{accents}
\usepackage[utf8]{inputenc}
\usepackage[T1]{fontenc}
\usepackage[english]{babel}
\usepackage{hyperref}
\usepackage{mathbbol,bbm}
\usepackage{float}

\let\originalleft\left
\let\originalright\right
\renewcommand{\left}{\mathopen{}\mathclose\bgroup\originalleft}
\renewcommand{\right}{\aftergroup\egroup\originalright}

\newcommand{\ket}[1]{\ensuremath{\left|#1\right\rangle}}

\newcommand{\ketbra}[2]{\ensuremath{\left| #1\right\rangle\left\langle#2\right|}}
\newcommand{\matelem}[3]{\ensuremath{\left\langle #1\middle|#2\middle| #3\right\rangle}}
\newcommand{\matelems}[3]{\ensuremath{\langle #1|#2| #3\rangle}}
\providecommand{\abs}[1]{\left|#1\right|}

\newcommand{\Tr}[1]{\ensuremath{\text{Tr}\left[#1\right]}}

\def\H{\ensuremath{H}}
\newcommand{\avg}[1]{\ensuremath{\overline{#1}}}
\newcommand{\ens}[1]{\ensuremath{\left\{#1\right\}}}
\newcommand{\enslabel}{\ensuremath{\lambda}}
\newcommand{\p}{\ensuremath{p}}
\newcommand{\dm}{\ensuremath{\rho}}
\newcommand{\avgdm}{\ensuremath{\avg{\rho}}}
\newcommand{\tmap}{\ensuremath{\Lambda}}
\newcommand{\id}{\mathbb{1}}
\newcommand{\Lop}{\ensuremath{L}}
\newcommand{\rate}{\ensuremath{\gamma}}

\newcommand{\qmemat}{\ensuremath{Q}}
\newcommand{\dynmat}{\ensuremath{\avg{F}}}

\newcommand{\define}{\ensuremath{\equiv}}
\newcommand{\gm}{\ensuremath{G}}
\newcommand{\Deltaz}{\ensuremath{\omega_0}}
\newcommand{\A}{\ensuremath{A}}
\newcommand{\f}{\ensuremath{\phi}}
\def\Im{\text{Im}}
\def\Re{\text{Re}}
\newcommand{\pwidth}{\ensuremath{\sigma}}
\newcommand{\Cop}{\ensuremath{C}}
\newcommand{\Cred}{\ensuremath{\tilde{C}}}
\newcommand{\frnd}{\ensuremath{\avg{\chi}}}
\newcommand{\sinc}{\text{sinc}}

\begin{document}

\title{Effective Dynamics of Disordered Quantum Systems}

\author{Chahan M. Kropf} 
\email{Chahan.Kropf@physik.uni-freiburg.de}
\affiliation{Physikalisches Institut, Albert-Ludwigs-Universit\"at Freiburg, Hermann-Herder-Stra{\ss}e~3, D-79104 Freiburg, Germany}

\author{Clemens Gneiting}
\affiliation{Physikalisches Institut, Albert-Ludwigs-Universit\"at Freiburg, Hermann-Herder-Stra{\ss}e~3, D-79104 Freiburg, Germany}

\author{Andreas Buchleitner}
\email{A.Buchleitner@physik.uni-freiburg.de}
\affiliation{Physikalisches Institut, Albert-Ludwigs-Universit\"at Freiburg, Hermann-Herder-Stra{\ss}e~3, D-79104 Freiburg, Germany}

\date{\today}

\begin{abstract}
\noindent
We derive general evolution equations describing the ensemble-average quantum dynamics generated by disordered Hamiltonians. The disorder average affects the coherence of the evolution and can be accounted for by suitably tailored effective coupling agents and associated rates which encode the specific statistical properties of the Hamiltonian's eigenvectors and eigenvalues, respectively. Spectral disorder and isotropically disordered eigenvector distributions are considered as paradigmatic test cases.
\end{abstract}

\maketitle


\section{Introduction}

{\em Disorder} is the expression of a lack of knowledge -- e.g., on a physical system's conformation or of a potential landscape, on scales that affect the system evolution. The characterization of a disordered system therefore requires a statistical approach, since reliable predictions on reproducible features of the system behavior can only be extracted by averaging over suitably chosen distributions of those uncontrolled properties. The specific choice of the distribution to be averaged over has potentially strong impact on the predicted behavior. In turn, the observed behavior may provide strong, though not necessarily unambiguous, hints on the underlying disorder's structure, 
as well known, e.g., from the classical Hamiltonian flow in mixed phase 
spaces (where one seeks a statistical characterization of the time evolution of ensembles of initial conditions) \cite{Mackay1984,Chirikov1984,Lai1992}.

On the quantum level, the disorder average has yet another important consequence: It implies an average over the accumulated phases associated with the eigenstates of every realization of the underlying random Hamiltonian (see Fig.~1). In general, this induces a loss of phase information, hence {\em decoherence}, while, simultaneously, dramatic disorder-induced interference effects may prevail under the disorder average. Note that, here, decoherence is an immediate consequence of the disorder average, and must not be mistaken for an irremediable loss of information to a large environment (bath) as in the context of open system. The  "lost" information is encoded in higher-order correlations such as intensity correlations of a speckle potential \cite{Goodman2007}, or in fluctuations of the measured conductance across individual samples \cite{Miniatura2011}. Such intricate indicators of coherence effects may, however, be difficult or even impossible to measure if single realizations of the disordered potential cannot be reliably resolved for several successive experimental runs as needed for the measurement of an observable. As an illustration, think of the propagation of photons in the presence of a turbulent atmosphere \cite{Herbst2015}. It therefore arises as a natural question what we can learn on the underlying disorder from the time-evolution of the ensemble averaged state alone, and especially from the coherent and incoherent content of the latter.

Prominent examples of robust interference effects which prevail even under the ensemble average are localization phenomena in the quantum transport theory of disordered or chaotic, dynamical systems \cite{kramer93,modugno10,wimberger_book,wellens09}. 
So far, however, these are discussed in terms of the associated spectral properties or of the signature of quantum interference in characteristic, experimentally observable quantities -- contrasted against classical predictions. Scattering theoretical approaches account for the above phase average and identify robust interference contributions, though almost always in the stationary state. Furthermore,  they only rarely quantify the relative weight of those coherent contributions which survive as compared to those which are eliminated under the disorder average, let alone the dynamical evolution of this ratio until stationarity is reached \cite{Cherroret2011b,Geiger2013a}. 

No formalism is so far available which allows to assess the effective dynamical 
quantum evolution of an arbitrary initial state in real time, and to directly associate characteristic time scales and 
couplings thereof with the underlying disorder. Note, however, that this represents a natural and -- given the wealth of disordered quantum 
transport problems -- substantial expansion of the generic playground
of the theory of open quantum systems, by substituting an uncontrolled environment by a static, operator-valued random perturbation  
of the system Hamiltonian. The relevance of such generalization of open system theory appears highly plausible, since it will pave the way for 
a systematic assessment of the question which type of disorder allows to exploit which type of quantum coherent phenomena, on 
transient and/or asymptotic time scales. Indeed, the direct experimental monitoring of the dynamical evolution of disordered quantum systems has now moved into reach \cite{Storzer2006,Semeghini2014,Muskens2012,Jendrzejewski2012}, and it therefore appears timely to develop tools to distinguish coherent and incoherent \cite{Cherroret2011b,Geiger2013a} features thereof. 

Here, we derive effective dynamical evolution equations for the ensemble-averaged state of disordered quantum systems, in the form of master equations, which, by their very structure, precisely meet the above purpose, and show how the statistics of the disorder enters the unitary part as well as the Lindblad operators and associated rates, as the equation's specific ingredients. We apply this theory to exemplary cases of quantum systems with random spectra and randomly distributed eigenvectors for which our method can be applied without any approximations. 

\begin{figure}[]
\includegraphics[width=\columnwidth]{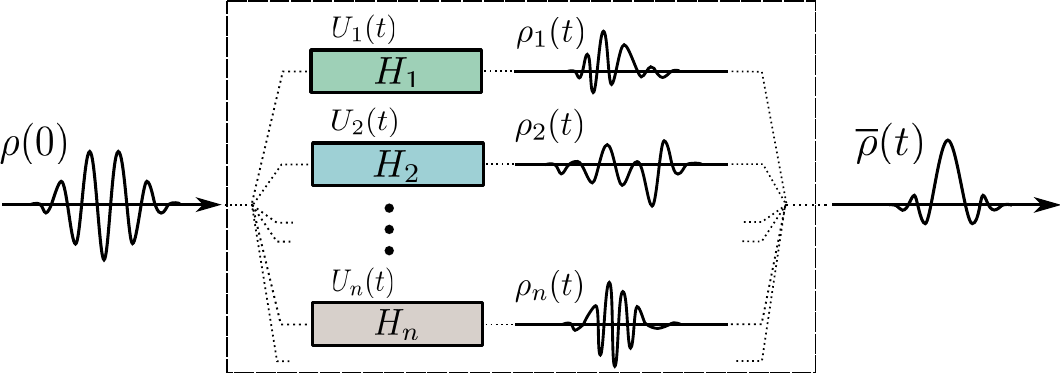}
	 \caption{[Color online]
Effective dynamics of a disordered quantum system. Different realizations $H_i$, $i=1,...n$, of the disordered system's Hamiltonian induce unitary time evolutions of the same initial state $\dm_0$ into distinct final states $\dm_i(t)=U_i(t) \dm_0 U_i^{\dagger}(t)$. The ensemble averaged state $\avgdm(t)$ is then obtained by 
averaging over all $\dm_i(t)$. The different realizations of the generating Hamiltonian may, e.g., result from slow parameter drifts between subsequent runs of an experiment, or from microscopically distinct potential landscapes in macroscopically identically prepared experimental settings.}
\label{fig:disorder-illustration}
\end{figure}

The article is structured as follows: In Section \ref{sec:dynamical_ensemble_average}, we introduce disordered quantum systems and the dynamics which emerge from the ensemble average. 
In Section \ref{sec:short-time_master_equation}, a direct derivation of a master equation valid in the limit of short times is presented (see also \cite{Gneiting2016}). Section \ref{sec:methodology} presents a more general method to obtain disorder master equations at all times. We then elaborate these methods for a single qubit with spectral disorder in Section \ref{sec:diagdisorder}, investigating the impact of different eigenvalue distributions. We continue in Section \ref{sec:diaognal-general-disorder} with the generalization of spectral disorder to higher dimensions and study the role of correlations among eigenvalues. Finally, we describe in Section \ref{sec:random matrics} the resulting master equations for unitarily invariant disorder ensembles, with which we illustrate the effect of disorder in the eigenvectors, and evaluate them for Gaussian and Poissonian eigenvalue statistics. Section \ref{sec:conclusions} concludes the paper.


\section{Disorder ensemble average} \label{sec:dynamical_ensemble_average}

To start with, let us briefly introduce quantum systems with disorder on the level of the Hamiltonian, and the corresponding ensemble average dynamics. Basic properties of the latter are exposed and their description in terms of quantum master equations outlined.

\subsection{Single realizations and ensemble average}

We consider an isolated, disordered quantum system of dimension $d$. The disorder may be characterized by an ensemble of time-independent Hamiltonians $\H_\enslabel$, occurring with probability $p_\enslabel$,
\begin{align} \label{eq:disorder_ensemble}
	&\ens{\left(\H_\enslabel,\; \p_\enslabel\right)},
\end{align}
where the (multi-)index $\lambda$ labels the different realizations. For each realization $\H_\enslabel$ of the disorder, the corresponding state $\dm_\enslabel(t)$ follows the von Neumann equation of motion,
\begin{equation} \label{eq:vonNeumann}
\dot \dm_\enslabel  (t) = -\frac{\rm i}{\hbar} \left[\H_\enslabel ,\dm_\enslabel  (t)\right],
\end{equation}
where $\dot{\dm}_\enslabel (t) = (d/dt) (\dm_\enslabel(t))$. Its solution is given by $\dm_\enslabel (t) = U_\enslabel (t)\dm(0)U_\enslabel ^\dagger(t)$, with the unitary time-evolution operator $U_\enslabel(t) = \text{exp}\left[-(i/\hbar) \H_\enslabel t\right]$. The initial ($t=0$) state $\rho(0)$ is taken to be identical for all realizations. 

The ensemble average state $\avgdm(t)$ (all ensemble average quantities will be marked with a bar) is obtained by the weighted sum over all realizations (for convenience denoted by an integral throughout),
\begin{align}
	&\avgdm(t) \define \int  d\enslabel \; \p_\enslabel\; \dm_\enslabel (t) = \int  d\enslabel \; \p_\enslabel U_\enslabel (t)\dm(0)U_\enslabel ^\dagger(t) .\label{eq:ensembleaverage}
\end{align}
For a given observable $B=\sum_b b \, \ketbra{b}{b}$, the ensemble average state delivers the average probability of measurement outcome $b$, which is given by $ \avg{p(b)} =\int d\enslabel p_\enslabel\,\Tr{\ketbra{b}{b} \dm_\enslabel} $, directly as
\begin{align}
	\avg{p(b)} = \Tr{\ketbra{b}{b} \avgdm (t)}. 
\end{align}
Formally, the dynamics of the ensemble average state (\ref{eq:ensembleaverage}) are described by a family of linear maps $\tmap_t$, parametrized by the time $t$, from the set of density matrices $\mathcal{D}$ onto itself:
\begin{align} \label{eq:tmap}
\tmap_{t}: \mathcal{D} &\rightarrow \mathcal{D} \\
\dm(0) &\rightarrow \avgdm(t) = \tmap_t\left[\dm(0)\right] \nonumber
\end{align}
In the following, we will describe some characteristic properties of this family of maps.

\subsection{Properties of the dynamical map}

The dynamics of each single realization is unitary and thus Hermiticity-preserving, trace-preserving, and completely positive. The linearity of the averaging procedure (\ref{eq:ensembleaverage}) then implies that $\tmap_t$ also has the latter three properties. Hence, the ensemble average dynamics describe by construction legitimate quantum dynamics \cite{Bengtsson2006}. But, in contrast to single realizations, the ensemble average dynamics are in general nonunitary (as we will explicitly show below), implying that Eq.~(\ref{eq:ensembleaverage}) cannot be subsumed by a single unitary operator.

A general property of the dynamical map is that the maximally mixed state is an invariant, 
$\tmap_t\left[\id\right] = \id$, indicating that the map is unital or bistochastic, in mathematical and statistical physics). In that sense the dynamics emerging from ensemble averages are more restricted than those of open systems, which can in general also be non-unital, as, e.g., in spontaneous decay processes of atoms.

As a direct consequence of the non-unitarity, purity, $\Tr{\rho^2}$, is in general not conserved. More precisely, the unital map $\tmap_t\left[\dm(0)\right]$ is always purity-decreasing, $\Tr{\dm^2(t)} \leq \Tr{\dm^2(0)}$ \cite{Streater1995}. As we will see below, this does however not imply that purity decays monotonously; it can increase locally, although it can never exceed the purity of the initial state.

\subsection{Master equation description}

For formal clarity, we here summarize several key properties of quantum master equations in general \cite{Breuer2002}, and especially in the context of disorder dynamics. 

As was briefly outlined in the introduction, we wish to describe the dynamics of the disorder ensemble average, formally defined by the map (\ref{eq:tmap}), in terms of a master equation. Since the ensemble average dynamics in general exhibit decoherence, they 
cannot be captured solely by the von Neumann equation of motion (\ref{eq:vonNeumann}) with some appropriately chosen Hamiltonian. Quantum master equations generalize the latter towards non-unitary dynamics while still guaranteeing the preservation of the Hermiticity and the trace of the state, and maintaining the complete positivity and the linearity of the underlying dynamical map. The general structure of quantum master equations imposed by these properties is given by the Lindblad form \cite{Lindblad1976,Gorini1976} 
\begin{align} \label{eq:qme}
&\dot{\avgdm}(t) = -\frac{i}{\hbar} \left[\mathcal{H}(t),\avgdm(t)\right] \\
& + \sum_k \rate_k(t) \left(\Lop_k(t)\avgdm(t) \Lop_k^\dagger (t) - \frac{1}{2} \left\{ \Lop_k^\dagger (t)\Lop_k(t), \avgdm (t) \right\}\right) . \nonumber 
\end{align}
This form complements the coherent dynamics of the von Neumann commutator (\ref{eq:vonNeumann}) by the incoherent dynamics induced by the Lindblad operators $\Lop_k(t)$ and their corresponding decoherence rates $\gamma_k(t)$. The curly brackets $\left\{A,B\right\} = AB+BA$ denote the anti-commutator. We remark that the Lindblad form (\ref{eq:qme}) does not uniquely fix the Hamiltonian and the set of Lindblad operators \cite{Breuer2002}. 
Let us stress that any integro-differential master equation including a memory kernel can be cast into the time-local form (\ref{eq:qme}) \cite{Chruscinski2010} (see also Appendix \ref{app:qme equivalence}). Possible memory-effects in (\ref{eq:qme}) are encoded in the time-dependence of the rates and Lindblad operators. We focus on this time-local form because it offers the physically most transparent interpretation for our purposes.

Our goal is to determine the relations between the composition of the disorder ensemble (\ref{eq:disorder_ensemble}) and the corresponding emerging master equation (\ref{eq:qme}), where the Lindblad operators and their rates will capture the collective (incoherent) dynamical impact of the disorder distribution. As we will show below, the ensemble average generically gives rise to a time-dependent Lindblad term, including negative decoherence rates (which are often considered as a signature of memory effects in the open system context \cite{Breuer2015}). While the latter can in general give rise to unphysical dynamics, in the considered case the emerging master equation will, by its very construction from the ensemble average, always be physically consistent. In the case of time-independent Lindblad operators and time-independent, positive rates, one usually speaks of a Markovian Lindblad master equation in the strict sense, describing a dynamical semi-group.


\section{Short-time master equation} \label{sec:short-time_master_equation}

For short times one can derive a general expression for the ensemble average master equation. To see this, let us consider the time evolution of a single realization over a time step $dt$, $\dm_\enslabel (dt) = U_\enslabel (dt)\dm(0)U_\enslabel ^\dagger(dt)$, and expand the time evolution operators to second order. One then obtains
\begin{align} \label{eq:short_time_single_realization}
\rho_{\lambda}(dt) =& ~\rho(0) + \frac{\rm i}{\hbar} dt [\rho(0), H_{\lambda}] \\
 & + \frac{dt^2}{\hbar^2} \left( H_{\lambda} \rho(0) H_{\lambda} - \frac{1}{2} H_{\lambda}^2 \rho(0) - \frac{1}{2} \rho(0) H_{\lambda}^2 \right) \nonumber \\
 & + \mathcal{O}(dt^3). \nonumber
\end{align}
As we will see, the expansion to second order is necessary, since the leading incoherent dynamical contributions only appear at this order. If we now take the ensemble average of (\ref{eq:short_time_single_realization}), isolating the contribution of the average Hamiltonian $\overline{H}$ from the second order term, we obtain
\begin{align} \label{eq:short_time_expansion}
\overline{\rho}(dt) =& ~\rho(0) + \frac{\rm i}{\hbar} dt [\rho(0), \overline{H}] \\
 & + \frac{dt^2}{\hbar^2} \left(\overline{H} \rho(0) \overline{H} - \frac{1}{2} \overline{H}^2 \rho(0) - \frac{1}{2} \rho(0) \overline{H}^2 \right) \nonumber \\
 & + dt^2 \int d\lambda \frac{p_{\lambda}}{\hbar^2} \left( L_{\lambda} \rho(0) L_{\lambda} - \frac{1}{2} L_{\lambda}^2 \rho(0) - \frac{1}{2} \rho(0) L_{\lambda}^2 \right) \nonumber \\
 & + \mathcal{O}(dt^3), \nonumber
\end{align}
where we have introduced the Hermitian operators $L_{\lambda} = H_{\lambda}-\overline{H}$. Eq.~(\ref{eq:short_time_expansion}) can be conceived as the second-order-in-time expansion of a master equation in Lindblad form (\ref{eq:qme}), which approximates the dynamics at short times 
(the step from Eq.~(\ref{eq:short_time_expansion}) to Eq.~(\ref{eq:short-time_master_equation}) is not completely trivial -- for more details 
see  \cite{Gneiting2016}):
\begin{align} \label{eq:short-time_master_equation}
\dot{\overline{\rho}}(t) =& - \frac{\rm i}{\hbar} [\overline{H}, \overline{\rho}(t)] \\
 & + \int d\lambda \gamma_{\lambda}(t) \left( L_{\lambda} \overline{\rho}(t) L_{\lambda}^{\dagger} - \frac{1}{2} \left\{ L_{\lambda}^{\dagger} L_{\lambda}, \overline{\rho}(t)\right\} \right) \nonumber
\end{align}
Each realization $\lambda$ gives in general rise to a Lindblad operator $L_{\lambda}$ and a corresponding decoherence rate $\gamma_{\lambda}(t)$,
\begin{equation} \label{eq:short-time_Lindbladians}
L_{\lambda} = \frac{H_{\lambda}-\overline{H}}{\hbar\omega_0} \qquad ; \qquad \gamma_{\lambda}(t) = 2 p_{\lambda} \omega_0^2t ,
\end{equation}
where $\hbar\omega_0$ is introduced as a characteristic energy scale of the system. We thus find that the ensemble average dynamics are described, at short times, by time-independent Lindblad operators given by the deviations of the single realization Hamiltonians from the average Hamiltonian, and by linearly-in-time increasing decoherence rates. Note that the above derivation is based on the second-order in time expansion (\ref{eq:short_time_single_realization})  and does not involve any further approximation, e.g., on the level of the ensemble averaging.

As we will see below, the particular structure of the disorder (e.g., when garnished by symmetries) will often allow us to transform (\ref{eq:short-time_master_equation}) into a 
significantly simplified master equation. Moreover, we will see that, in our cases, the short-time master equation already reflects the structure of the master equation at arbitrary times, with only the time dependence of the decoherence rates modified, while in other cases the finite-time dynamics may also give rise to modified Lindblad operators. The coherent part of the short-time dynamics is induced by the average Hamiltonian $\overline{H}$. Note that, since the decoherence rate vanishes at $t=0$, the Lindblad term (the second line in Eq.~(\ref{eq:short-time_master_equation})) does not contribute to first order in time. Furthermore, we remark that in deriving the short-time master equation (\ref{eq:short-time_master_equation}) we had to assume that the average Hamiltonian $\overline{H}$ exists; if this is not the case, the short-time dynamics may differ from (\ref{eq:short-time_master_equation}, \ref{eq:short-time_Lindbladians}), as in the examples which will be discussed in Sections~\ref{Sec:Cauchy-Lorentz} and \ref{Sec:Levi} below.


\section{Finite-time master equation} \label{sec:methodology}

We are not aware of a general expression, similar to the short-time master equation (\ref{eq:short-time_master_equation}), for the ensemble average master equation at arbitrary times. In this section, we outline a method that can be employed in order to determine the master equation for arbitrary disorder distributions given the following holds: Let us assume that the inverse $\tmap^{-1}_{t}[\rho]$ of the dynamical map $\tmap_t\left[\dm\right]$ defined in Eqs.~(\ref{eq:ensembleaverage}) and (\ref{eq:tmap}) exists. We can then write $\dm(0) = \tmap^{-1}_{t}\left[\avgdm(t)\right]$. A time-local differential equation is formally obtained by taking the time derivative of Eq.~(\ref{eq:tmap}),
\begin{align}\label{eq:formal qme}
		\dot{\avgdm}(t) & = \dot{\tmap}_{t}\left[\avgdm(0)\right] = \dot{\tmap}_{t}\circ\tmap^{-1}_t \;\left[\avgdm(t)\right],
\end{align}
where $\circ$ denotes map composition. Note that this can be done for any sufficiently well-behaved dynamical map $\tmap_t\left[\dm\right]$. The issue of the possible non-existence of the inverse $\tmap^{-1}_{t}[\rho]$ will be discussed at the end of this Section.

In the following we describe an explicit method to compute the maps $\dot{\tmap}_{t}$ and $\tmap^{-1}_t$, and how to obtain from this a master equation in Lindblad form. The method is based upon a matrix approach presented by Andersson and Hall \cite{Andersson2007, Hall2014}. It can in principle be applied to any kind of Hamiltonian disorder, but the computational complexity can be substantial, making approximations often unavoidable. Note, however, that for all the examples presented in this paper (see Sections \ref{sec:diagdisorder}-\ref{sec:random matrics}), we can derive the disorder master equation without any approximations.

Expressed in terms of an orthonormal basis $\ens{\ket{j}}_{j=1}^{d}$, the ensemble average state (\ref{eq:ensembleaverage}) reads
\begin{align}
	\matelem{j}{\avgdm(t)}{k} = \sum_{r,s=1}^d \;\matelem{r}{\dm(0)}{s} \;\int d\enslabel \, p_\enslabel  \matelem{j}{U_{\enslabel}(t)}{r} \matelems{s}{U_{\enslabel} ^\dagger(t)}{k} .
\end{align}
The basis $\ens{\ket{j}}_{j=1}^{d}$ can be chosen suitably to ease the computation of the inverse of the average dynamical matrix defined in Eq.~(\ref{eq:dynamical matrix}) below. For convenience we adopt a vector notation and define the $d^2\times d^2$ average dynamical matrix $\dynmat (t)$ and the $d^2\times1$ average density vector $\vec{\avgdm}$ component-wise by
\begin{align} \label{eq:dynamical matrix}
	\dynmat_{jk,rs}(t) \define \int d\enslabel \, p_\enslabel \matelem{j}{U_{\enslabel}(t)}{r}\matelems{s}{U_{\enslabel} ^\dagger(t)}{k}
\end{align}
and $\vec{\avgdm}_{jk}(t) \define \matelem{j}{\avgdm(t)}{k}$, where $(jk)$ and $(rs)$ are double indices with $j,k,r,s \in \left\{1,2,...d\right\}$. We remark that the average dynamical matrix $\dynmat(t)$ contains the same information as the $d^2\times d^2$ Choi matrix \cite{Choi1975}. (Equivalent dynamical matrices with different index orderings exist in the literature \cite{Sudarshan1961,Sudarshan2003}.)

In terms of the average dynamical matrix $\dynmat (t)$ and the average density vector $\vec{\avgdm}$, the ensemble average state (\ref{eq:ensembleaverage}) is obtained by the standard matrix product $\vec{\avgdm}(t) = \dynmat(t) \cdot \vec{\dm}(0)$. Based on this representation, the inverse and the time derivative of the dynamical map can be computed using standard matrix operations. Concretely, the differential equation (\ref{eq:formal qme}) can now be written as
\begin{align}
	\dot{\vec{\avgdm}}(t) = \dot{\dynmat}(t) \cdot \dynmat^{-1}(t) \cdot \vec{\avgdm} (t) = \qmemat(t) \cdot \vec{\avgdm}(t) , \label{eq:matrixqme}
\end{align}
where the $d^2\times d^2$ matrix $\qmemat(t) \define \dot{\dynmat}(t) \cdot \dynmat^{-1}(t)$ 
represents the map $\dot{\tmap}_t\circ \tmap_t^{-1}$. This implies, in terms of the components of $\qmemat$ and $\vec{\avgdm}$:
\begin{equation}	
	\dot{\avgdm}(t) = \sum_{j,k,r,s=1}^d \qmemat (t)_{jk,rs} \; \ketbra{j}{r} \avgdm (t) \ketbra{s}{k} . \label{eq:qmewithG}
\end{equation}
The final Lindblad form (\ref{eq:qme}) is then obtained by expanding the operators $\ketbra{j}{r}$ and $\ketbra{s}{k} $ in a Hermitian operator basis and collecting the different terms using the hermicity of $\avgdm(t)$ and trace-preservation, $\text{Tr}\left[\dot{\avgdm}(t)\right]=0$ (this step is also performed in the textbook derivation of the Lindblad master equation in open quantum systems theory \cite{Breuer2002}) . More precisely, one chooses a Hermitian operator basis $\left\{\A_m\right\}_{m=0}^{d^2-1}$ with 
\begin{align}\label{eq:A basis}
\A_0 = \frac{1}{\sqrt{d}}\id \;\; ; \;\; \A_m = \A_m^\dagger \;\; ; \;\; \text{Tr}\left[\A_m\A_n\right]=\delta_{mn}.
\end{align}
By setting $n=0$, the orthogonality relation implies $\text{Tr}\left[\A_m\right]|_{m\neq 0}=0$. Any operator basis satsifying (\ref{eq:A basis}) can be chosen (note that the Hermicity condition is for convenience only). However, a suitable choice may simplify subsequent calculations. A natural choice are, e.g., the $d^2-1$ Gell-Mann matrices \cite{Bertlmann2008},  which are a direct extension of the Pauli matrices to higher dimensions.

The Gell-Mann matrices (here denoted with $\gm$ in order to avoid confusion with the unspecified basis operators $A$) can be separated into sets of symmetrical $\ens{\gm_s}$, antisymmetrical $\ens{\gm_a}$, and diagonal $\ens{\gm_d}$ matrices. Given the basis $\ens{\ket{j}}_{j=1}^d$, we have, with the normalization required by (\ref{eq:A basis}),
\begin{align}\label{eq:Gell-Mann-matrices}
	\gm_s^{jk} &= \frac{1}{\sqrt{2}}(\ketbra{j}{k} + \ketbra{k}{j}) \;\; \text{for} \;\; 1\leq j < k \leq d, \\
	\gm_a^{jk} &= \frac{1}{\sqrt{2}}(-i\ketbra{j}{k} + i\ketbra{k}{j}) \;\; \text{for} \;\; 1\leq j < k \leq d, \nonumber \\
	\gm_d^{l} &= (\frac{1}{\sqrt{l(l+1)}})\left(-l\ketbra{l+1}{l+1}
	 +\sum_{j=1}^l \ketbra{j}{j}\right) \nonumber \\
	 &\; \hspace{4cm} \text{for} \;\; 1\leq l \leq d-1. \nonumber
\end{align}

Coming back to the general operator basis (\ref{eq:A basis}), Eq.~(\ref{eq:qmewithG}) rewritten in terms of the $A_m$ yields (here and in the following we drop the time-dependence of $\avgdm$ in the notation of the master equations)
\begin{align} \label{eq:CtildeQme}
	\dot{\avgdm} = \Cred (t)\avgdm+\avgdm \, \Cred^\dagger(t) +\sum_{m,n=1}^{d^2-1} \Cop_{mn}(t)\A_m \avgdm \A_n ,
\end{align}
where we introduced for clarity, and as an intermediary step, the Hermitian matrix
\begin{align} \label{eq:C matrix}
	\Cop_{mn}(t) = \sum_{j,k,r,s=1}^d \qmemat (t)_{jk,rs} \; \A_{m, r j} \A_{n,k s}
\end{align}
with the abbreviation $\A_{m, r j} = \matelem{r}{\A_m}{j}$, the $rj$ component of $\A_m$.
The terms with $m=0$ or $n=0$ are separately collected in
\begin{align}
	\Cred(t) = \frac{\Cop_{00}(t)}{2d}\id + \sum_{m=1}^{d^2-1}\frac{\Cop_{m0}(t)}{\sqrt{d}}\A_m .
\end{align}
From Eq.~(\ref{eq:CtildeQme}) and since $\text{Tr}\left[\dot{\avgdm}\right]=0$, one finds that 
$\left( \Cred(t)+\Cred^\dagger(t) \right) = -\sum_{m,n=1}^{d^2-1} \Cop_{mn}(t)\A_m^\dagger \A_n $. If we now introduce the {\em effective} Hamiltonian
\begin{align} \label{eq:C-Hamiltonian}
H(t)=\frac{i\hbar}{2} \left( \Cred(t)-\Cred^\dagger(t) \right),
\end{align}
we arrive at the non-diagonal Lindblad form
\begin{align} \label{eq:non-diagonal_Lindblad_form}
	 	\dot{\avgdm}(t) = &-\frac{i}{\hbar} \left[\H(t),\avgdm(t)\right] \\&+ \sum_{m,n=1}^{d^2-1} \Gamma_{m n}(t) \left(\A_{m}\avgdm(t) \A_{n}^\dagger - \frac{1}{2} \left\{\A_n^\dagger \A_m, \avgdm (t) \right\}\right) , \nonumber
\end{align}
with the decoherence matrix $\Gamma_{m n}(t) = \Cop_{mn}(t)$, ($m,n>0$, i.e., the matrix (\ref{eq:C matrix}) without the first line and the first column), 
and the (Hermitian, time-independent) Lindblad operators $\A_m$.

Note that the ensemble average (\ref{eq:ensembleaverage}), even when taken over the unitary dynamics arising from static Hamiltonians as considered in this article, gives rise to a possibly time-dependent effective Hamiltonian $\H (t)$, which does in general not coincide with the average Hamiltonian $\avg{\H}$. We emphasize that this time-dependence does \emph{not} mediate the effect of an external driving potential as would be usual in the context of open quantum systems, but is a direct consequence of the composition of the disorder ensemble (\ref{eq:disorder_ensemble}). An example is discussed in detail in Section \ref{Sec:Levi}.

Since the decoherence matrix is Hermitian, $\Gamma(t)=\Gamma^\dagger(t)$, it can be diagonalized, $\Gamma_{mn}(t) = \sum_{k=1}^{d^2-1} V_{mk} (t) \gamma_k(t) V^\dagger_{kn}(t)$, with $V_{mk}(t)$ and $\gamma_k(t)$ the $k$th eigenvector and eigenvalue of $\Gamma(t)$, respectively. This step requires explicit knowledge of the decoherence matrix $\Gamma(t)$. We then obtain the diagonal Lindblad form (\ref{eq:qme}) of the disorder master equation with Lindblad operators $\Lop_k(t)=\sum_{m=1}^{d^2-1} V_{mk} (t) \A_m$ and strictly real decay rates $\gamma_k(t)$. In general, the diagonalization leads to time-dependent and non-Hermitian Lindblad operators. However, below we will also give an example of a disorder distribution giving rise to a time-independent master equation.

Note that, in the spirit of the short-time expansion in Sec.~\ref{sec:short-time_master_equation}, one can expand the decoherence matrix $\Gamma(t)$ in time, $\Gamma_{m n}(t) = \Gamma_{m n}(0)+\dot \Gamma_{m n}(0) t+ \ddot \Gamma_{m n}(0) t^2/2+\dots$ This then reexpresses the nondiagonal Lindblad master equation, where each order of the time expansion of $\Gamma(t)$ leads to a Lindblad term, i.e.,~each time order can be diagonalized independently. 
When all derivatives $\frac{d^k}{dt^k} \Gamma_{m n}(0)$ commute, there exists a set of time-independent Lindblad operators specifying the diagonal form (\ref{eq:qme}) at all times. If one truncates the expansion after the linear term, one recovers the incoherent part of the short-time master equation (\ref{eq:short-time_master_equation}) (provided it exists, i.e., provided $\avg{\H}$ is well defined, which is not the case for certain disorder distributions, such as for example the Cauchy-Lorentz diagonal disorder considered in \ref{Sec:Cauchy-Lorentz}), however in a different representation. 

The presented method to convert a dynamical map into a time-local master equation requires the existence of the inverse map $\Lambda_t^{-1}[\rho]$, cf.~Eq.~(\ref{eq:formal qme}). A necessary condition for the inverse not to exist is that two different initial states evolve into the same state at a given finite time $t'>0$ \cite{Hall2014}. As we will see below (e.g.~in~Section \ref{sec:diagdisorder}), such coincidence can indeed occur for the ensemble average dynamics. However, this does not necessarily imply that one cannot find the corresponding master equation. If the inverse does not exist only at isolated points in time, one can still formulate a master equation which then reflects the non-existence of the inverse by diverging decoherence rates. In Section \ref{sec:diagdisorder}, diverging rates will for example arise for a single qubit subject to uniform spectral disorder. As we will see, there the divergences are a consequence of the compact support of the disorder distribution.

Indeed, in all situations considered in this article, the inverse map exhibits at most isolated divergences. We conjecture that this is a general property of dynamical ensemble averages on finite-dimensional Hilbert spaces as defined in (\ref{eq:ensembleaverage}),
due to the quasiperiodicity of the time-evolved state inherited from the discrete spectrum of $U_\lambda$.

Finally, we emphasize that the method to derive a master equation starting from a quantum dynamical map $\Lambda_t[\rho]$ presented in this section is not restricted to ensemble average dynamics, but can be applied to any invertible dynamical map. Note however that, in some cases, other, more direct, methods, may be preferable. For instance, in the case of unitarily invariant disorder (cf.~Section \ref{sec:random matrics}), or when considering random unitary channels as they are usually defined in quantum information theory ($\avgdm = \sum_j p_j(t) U_j \dm U_j^\dagger$) \cite{Chruscinski2015}, one can directly invert the map $\tmap_t$. The Choi-Jamilkowsky isomorphism \cite{Jamiolkowski1972,Jiang2013} may provide us with yet another alternative method.


\section{Spectral disorder: single qubit}\label{sec:diagdisorder}

The first conceptual benchmark situation we focus on has the disorder occurring in the spectrum of the Hamiltonians, while all realizations share the same eigenvectors (for the moment we consider the general case of a $d$-dimensional quantum system),
\begin{align}\label{eq:diagonal-H}
\H_{\vec{\enslabel}} = \hbar \Deltaz \sum_{j=1}^d \enslabel_j \ketbra{j}{j}.
\end{align}
The disorder is then fully characterized by the distribution $p_{\vec{\enslabel}}$ of the eigenenergies $\hbar \Deltaz \vec{\enslabel} \define \hbar \Deltaz (\enslabel_1,...,\enslabel_d)^{T}$, where $\Deltaz \in \mathbb{R}^+$ is the characteristic Larmor precession frequency. Such 
scenario for example describes an ensemble of non-interacting spins $1/2$ in a static magnetic field with spatial inhomogeneities, or which fluctuates in intensity from one measurement to another, a situation for instance encountered in magnetic resonance spectroscopy \cite{Slichter1990} or in ion trap experiments \cite{Gessner2014}.

We start by considering the case of a spectrally disordered qubit (i.e.~a system of dimension $d=2$), illustrating in some detail the matrix approach to compute the disorder quantum master equation outlined in Section \ref{sec:methodology}. As we show, even for such a simple system, the ensemble average can exhibit nontrivial dynamics which become transparent on the level of the disorder quantum master equation.

We parametrize the random Hamiltonian ensemble by
\begin{align*}
	\ens{ \left( \H_\enslabel = \enslabel \hbar \frac{\Deltaz}{2}  \sigma_z \;\;,\;\;p_\enslabel = p(\enslabel) \right) } \;\;\; ; \;\; \sigma_z = \begin{pmatrix}
		1 & 0 \\ 0 & -1
	\end{pmatrix}
\end{align*}
with a single, dimensionless disorder parameter $\enslabel$, and the corresponding probability distribution $p(\enslabel)$. Note that, for simplicity and without loss of generality, the single parameter $\enslabel$ here describes, in contrast to (\ref{eq:diagonal-H}), the disorder in terms of the energy level difference. In the eigenbasis $\{\ket{1},\ket{2}\}$ of the Pauli matrix $\sigma_z$ and with the initial conditions $\dm_{jk}(0)\equiv\matelem{j}{\dm(0)}{k},\; j,k=1,2$, the ensemble average dynamics are given by
\begin{align}\label{eq:qubit rho(t)} 
	\avgdm(t) &=\begin{pmatrix}
		\dm_{11}(0) & \f^*(\Deltaz t)\dm_{12}(0) & \\
		\f(\Deltaz t)\dm_{21}(0) & \dm_{22}(0)
	\end{pmatrix} ,
\end{align}
where $\f^*$ denotes the complex conjugate of $\f$. The off-diagonal dynamics are captured by the characteristic function \cite{Lukacs1972} $\f(t)$ of the probability distribution $p(\enslabel)$,
\begin{align}
		\f(\Deltaz t) &\define \int_{-\infty}^{\infty} d\enslabel \, p(\enslabel)e^{i \enslabel \Deltaz t}.\label{eq:qubitf(t)}
\end{align}
Using (\ref{eq:qubit rho(t)}), we can immediately derive the dynamical matrix $\dynmat$ defined in Eq.~(\ref{eq:dynamical matrix}). With the indices ordered as (11), (12), (21), (22), such that $\vec{\avgdm}(t) = \left(\dm_{11}(0),\f(\Deltaz t)\dm_{12}(0),\f^*(\Deltaz t)\dm_{21}(0),\dm_{22}(0)\right)^T$, we get
\begin{align}
	\dynmat(t)=\begin{pmatrix}
		1 & 0 & 0 & 0 \\
		0 & \f^*(\Deltaz t) & 0 & 0 \\
		0 & 0 & \f(\Deltaz t) & 0 \\
		0 & 0 & 0 & 1
	\end{pmatrix} .
\end{align}
Time derivative and inverse of $\dynmat(t)$ are thus easily computed and the disorder master equation (\ref{eq:matrixqme}) is then determined by $\qmemat(t) = \dot{\dynmat}(t) \cdot \dynmat^{-1}(t)$, where we must restrict to times $t\geq 0$ (because the initial state is at $t=0$).

In order to obtain the desired Lindblad form, we use the Hermitian operator basis (\ref{eq:Gell-Mann-matrices}) with $\A_0=\id /\sqrt{2}$ and $\A_j=\sigma_j/\sqrt{2}$, with $j=1,2,3$ and $\sigma_j$ the Pauli matrices. We then obtain for the matrix $\Cop$ defined in (\ref{eq:C matrix})
\begin{align}
		\Cop(t)&=
		\begin{pmatrix}
		\Re{\frac{\dot{\f}(\Deltaz t)}{\f(\Deltaz t)}} & 0 & 0 & i\Im{\frac{\dot{\f}(\Deltaz t)}{\f(\Deltaz t)}} \\
		0 & 0 & 0 & 0 \\
		0 & 0 & 0 & 0 \\
		-i\Im{\frac{\dot{\f}(\Deltaz t)}{\f(\Deltaz t)}} & 0 & 0 & -\Re{\frac{\dot{\f}(\Deltaz t)}{\f(\Deltaz t)}}
	\end{pmatrix}.
\end{align}
The final form of the master equation follows directly from Eq.~(\ref{eq:C-Hamiltonian}), and noting that $\Cop_{m n}$ ($m,n > 0$) is already diagonal. One obtains
\begin{align} \label{eq:qubitqme}
	\dot{\avgdm} = -\frac{i}{\hbar}\left[ \varphi(t) \sigma_z,\avgdm \right] + \rate(t) \Bigl(\sigma_z \avgdm \sigma_z^\dagger -\avgdm \Bigr) ,
\end{align}
with the energy function
\begin{align} \label{eq:qubitphi}
	\varphi(t) &= \frac{\hbar}{2}\text{Im}\left[\frac{d}{dt} \ln \left(\f(\Deltaz t)\right)\right]
\end{align}
and the in general time-dependent decoherence rate
\begin{align} \label{eq:qubitgamma}
	\rate(t) & = -\frac{1}{2}\text{Re}\left[\frac{d}{dt} \ln \left(\f(\Deltaz t)\right)\right],
\end{align}
with $\ln(\cdot)$ the principal branch of the complex logarithm.
We thus obtain, for an arbitrary probability distribution $p(\enslabel)$, a dephasing master equation (\ref{eq:qubitqme}), where the energy function $\varphi(t)$ and the decoherence rate $\rate(t)$ depend on the specific choice of $p(\enslabel)$. By dephasing we understand that the diagonal elements of the density matrix are time-independent, while the off-diagonal elements evolve non-unitarily; here according to 
\begin{align}\label{eq:qubit-dynamics-solved}
 \avgdm_{12}(t)=\dm_{12}(0)e^{-2\int_0^t dt'  \left((i/\hbar)\varphi(t') + \gamma(t')\right)}.
\end{align}
In the simplest case, the off-diagonal elements will decay monotonously, but, as we will see in the examples below, revivals can also occur.

The master equation (\ref{eq:qubitqme}) thus comprehensively captures a dynamical effect which is familiar, for instance, from nuclear magnetic resonance experiments with spin 1/2 nuclei, where spatial inhomogeneities in the external magnetic field give rise to a distribution of precession frequencies. The ensemble average then amounts to averaging over these frequencies, resulting in the described dephasing given the frequencies are not all commensurate to one another. This dephasing is there characterized by the decay time $T_2$ \cite{Slichter1990}.

Alternatively to Eqs.~(\ref{eq:qubitphi}) and (\ref{eq:qubitgamma}), one can express the energy function $\varphi(t)$ and the decoherence rate $\gamma(t)$ in terms of the cumulants $\kappa^{(n)}$ of the characteristic function $\phi(\Deltaz t)$ (see also Appendix~\ref{app: characteristic fct}):
\begin{align}
	\varphi(t) &= \frac{\hbar}{2}\left(\Deltaz\kappa^{(1)} + \sum_{n=1}^\infty \frac{(-1)^n}{(2n)!} \; {\Deltaz}^{2n+1}\kappa^{(2n+1)}\; t^{2n} \right) \, ,\label{eq:qubitphi_cumulant} \\
	\rate(t) & = \frac{1}{2}\left(\Deltaz^2\kappa^{(2)}\;t-\sum_{n=2}^\infty \frac{(-1)^n}{(2n-1)!} \; {\Deltaz}^{2n}\kappa^{(2n)}\; t^{2n-1} \right) \, .\label{eq:qubitgamma_cumulant}
\end{align}
The energy function $\varphi(t)$ depends only on the odd cumulants, which encode the degree of (a-)symmetry in the distribution (e.g. $\kappa^{(1)}$ equals the average value, and $\kappa^{(3)}$ is proportional to the skewness, i.e., "degree of asymmetry", of the distribution). The decoherence rate $\gamma(t)$ is a function of only the even cumulants, which characterize the broadness of the distribution (e.g. $\kappa^{(2)}$ and $\kappa^{(4)}$ are proportional to the variance and to the kurtosis, i.e., "degree of peakedness", of the distribution, respectively). We thus learn that the coherent part of the dynamics is governed by the symmetry properties of the eigenvalue distribution, while the disorder broadness (or its strength) controls the incoherent dephasing dynamics. For example, if the disorder stems from uncontrolled parameter variations of experimental apparatuses over repeated measurements, its asymmetries would give rise to systematic deviations from the desired (coherent) dynamics, while the incurred statistical error results in the decoherence. 

For distributions $p(\enslabel)$ that are symmetric with respect to their average value, such as a Gaussian or a uniform distribution, odd cumulants of order $n>1$ vanish. In that case, the coherent part of the master equation (\ref{eq:qubitqme}) is driven by the average Hamiltonian $\avg{H}$. Irrespective of that, there will also be an incoherent contribution from the even cumulants. Therefore, the ensemble average dynamics does, also for symmetric distributions, not coincide with the unitary dynamics induced by the average Hamiltonian, i.e., $\avg{\exp(-i H_{\enslabel} t/\hbar) \rho_0 \exp(i H_{\enslabel} t/\hbar)} \neq \exp(-i \avg{H} t/\hbar) \rho_0 \exp(i \avg{H} t/\hbar)$.

Before evaluating the disorder-induced dephasing master equation (\ref{eq:qubitqme}) for various paradigmatic disorder distributions $p(\lambda)$, let us consider its short-time approximation. The leading-order expansions of the energy function $\varphi(t)$, Eq.~(\ref{eq:qubitphi_cumulant}), and the decoherence rate $\gamma(t)$, Eq.~(\ref{eq:qubitgamma_cumulant}), for short times $t\ll 1/\Deltaz$, yield
\begin{align} \label{eq:qubit_leading_order}
	\varphi(t) \approx \frac{\hbar}{2}\Deltaz \avg{\enslabel}\;\; ; \;\;
	\rate(t) \approx \frac{1}{2}\Deltaz \left(\avg{\enslabel^2}-\avg{\enslabel}^2\right)\;t ,
\end{align}
where we used that $\kappa^{(1)}=\avg{\enslabel}$ and $\kappa^{(2)} = \left(\avg{\enslabel^2}-\avg{\enslabel}^2\right)$. This can be compared to the short-time master equation (\ref{eq:short-time_master_equation}). With the average Hamiltonian $\avg{\H} = \avg{\enslabel}\hbar (\Deltaz /2) \sigma_z$, the short-time Lindblad operators and rates evaluate according to Eq.~(\ref{eq:short-time_Lindbladians}) as
\begin{align}
	L_\enslabel = \frac{1}{2}(\enslabel-\avg{\enslabel})\sigma_z \;\; ; \;\;
	\gamma_\enslabel(t) = 2p(\enslabel)\Deltaz^2 t.
\end{align}
One can now explicitly perform the disorder integral in Eq.~(\ref{eq:short-time_master_equation}), yielding the short-time master equation for qubit spectral disorder,
\begin{align} \label{eq:qubit short-time-diagonal-qme}
	\dot{\avgdm} = -\frac{i\Deltaz \avg{\enslabel}}{2} \; \left[\sigma_z,\avgdm \right] +\frac{\Deltaz^2 t}{2}\left(\avg{\enslabel^2}-\avg{\enslabel}^2\right) \; \Bigl(\sigma_z \avgdm \sigma_z^\dagger -  \avgdm\Bigr). 
\end{align}
As expected, this coincides with the leading-order expansion (\ref{eq:qubit_leading_order}) of the exact master equation (\ref{eq:qubitqme}), showing that the short-time master equation corresponds to the leading order of the cumulant expansion of the energy function and of the decoherence rate. 

In the following, we illustrate the variety of the single-qubit dephasing dynamics that can arise from spectral disorder with four paradigmatic disorder distributions $p(\enslabel)$: a Cauchy-Lorentz, a Gaussian, a uniform box, and a L\'{e}vy distribution. We will see that the distinctive properties of these distributions are clearly reflected in the temporal evolution of the associated respective decoherence rates. The examples are chosen to illustrate the range of possible relations between the properties of the disorder distribution and the master equation describing the resulting ensemble average dynamics. Apart from their paradigmatic relevance, these distributions also describe realistic physical situations. For instance, as argued above, spectral disorder can account for part of the inhomogeneous broadening of the lineshapes in spectroscopy experiments. In such a setting, the Cauchy-Lorentz distribution would be related to static field inhomogeneities \cite{Slichter1990}, and the Gaussian distribution to Doppler broadening or to the broadening due to impurities \cite{Stoneham1969,Siegman1986}.  A uniform box distribution is used to model a finite-bandwidth of the distribution. The L\'{e}vy distribution is characteristic for broad-band disorder and has been used to study the spectral lines of excitonic states in molecular systems or assemblies of Rydberg atoms \cite{Eisfeld2010,Barkai2000}.

\subsection{Cauchy-Lorentz distribution} \label{Sec:Cauchy-Lorentz}
We begin with a Cauchy-Lorentz distribution, which is familiar from resonance phenomena in the frequency domain and results in constant decay rates in the time domain, leading to an exponential decay behavior. As we will show, this expectation is confirmed in our context as well. We parametrize the Cauchy-Lorentz distribution $p_{\rm CL}(\enslabel)$ by
\begin{equation}
	p_{\rm CL}(\enslabel) = \frac{1}{\pi}\frac{\pwidth}{(\enslabel-\enslabel_0)^2+\pwidth^2}\;,\;\pwidth\in\mathbb{R}^+,\;\enslabel_0\in\mathbb{R} ,
\end{equation}
with the (dimensionless) scale parameter $\sigma$ and the (dimensionless) location parameter $\lambda_0$. According to (\ref{eq:qubitf(t)}), this gives rise to the characteristic function $\phi_{\rm CL}(\Deltaz t) = \exp[i\enslabel_0\Deltaz t-\pwidth\Deltaz t]$. The energy function (\ref{eq:qubitphi}) and the decoherence rate (\ref{eq:qubitgamma}) read
\begin{align} \label{eq:qubit-lorentz-rates}
	 \varphi_{\rm CL} = \hbar \frac{\Deltaz}{2} \enslabel_0 \;\; ; \;\; \rate_{\rm CL} = \frac{\Deltaz}{2} \pwidth .
\end{align}
As we can see, the location parameter $\enslabel_0$ specifies the time-independent energy function and, thus, fixes the Larmor frequency $\Deltaz\enslabel_0$ of the coherent precession about the $z$-axis induced by the Hamiltonian part of the master equation Eq.(\ref{eq:qubitqme}) (see also Fig.~\ref{fig:qubit-all-coherent} of Appendix \ref{app:qubit-coherent}). The disorder scale parameter $\pwidth$ sets the time-independent dephasing rate $\gamma_{\rm CL}$ and, thereby, determines the strength of the dephasing process.  More precisely, the time-independent rate $\gamma_{\rm CL}$ leads to a purely exponential decay with a rate proportional to $\pwidth$ of the off-diagonal terms of the density matrix (see Figure \ref{fig:qubit-all}), since solving the disorder master equation (\ref{eq:qubitqme}) yields $\matelem{1}{\avgdm(t)}{2} \propto \exp\left(-2\int_0^t \gamma(t')dt'\right) = \exp\left(-\abs{\Deltaz} \pwidth t\right)$ (c.f. Eq.(\ref{eq:qubit-dynamics-solved})). In nuclear magnetic resonance spectroscopy, $1/\gamma_{\rm CL}$ would characterize the contribution to the dephasing time $T_2^*$ coming from the magnetic field inhomogeneity. The Cauchy-Lorentz distribution is distinguished in that it is the only disorder distribution which gives rise to both a time-independent decoherence rate $\gamma_{CL}$ and a time-independent energy function $\varphi_{CL}$, resulting in a purely Markovian Lindblad master equation obeying the semi-group property \cite{Lindblad1976,Gorini1976}.

We note that all moments and cumulants $\kappa^{(n)}$ of the Cauchy-Lorentz distribution $p_{\rm CL}(\enslabel)$ are diverging. Nevertheless, the characteristic function $\phi_{\rm CL}(\Deltaz t)$ of $p_{\rm CL}(\enslabel)$ (i.e.~its Fourier transform) is well defined, as is it the case for any probability distribution, and consequently so are also the decoherence rate and the energy function, which are computed analytically. The energy function $\varphi_{\rm CL}$ is driven by the central Larmor frequency $\Deltaz \enslabel_0$ because of the symmetry of the Cauchy-Lorenty distribution.

As a consequence of the non-existing moments and cumulants of the Cauchy-Lorentz distribution, the emerging dynamics exhibit a short-time behavior which differs from the one described by the short-time master equation (\ref{eq:qubit short-time-diagonal-qme}), in that (\ref{eq:qubit short-time-diagonal-qme}) predicts wrongly (at short times) a linearly-in-time increasing decoherence rate instead of the correct time-independent $\gamma_{\rm CL}$ 
(such linear increase would induce a Gaussian rather than an exponential decay of the coherences). The correspondence to the short-time description (\ref{eq:qubit short-time-diagonal-qme}) can be restored by introducing a frequency cut-off that suppresses the algebraically decaying tails of $p_{\rm CL}(\enslabel)$, e.g.,~$\enslabel \in \left[-a,a\right], \; a>0$. This then corresponds to a more physical picture than the pure exponential decay, as can for example be seen in the initial Gaussian shape of the free induction decay signal in nuclear magnetic resonance experiments \cite{Slichter1990}.

\subsection{Gaussian distribution}
Next, we consider the ubiquitous Gaussian (or normal) distribution $p_{\rm G}(\enslabel)$, which, in contrast to many other contexts, gives here rise to intriguing consequences. It is defined as
\begin{equation} \label{eq:qubit-distribution-gaussian}
	p_{\rm G}(\enslabel) = \frac{1}{\pwidth\sqrt{2\pi}}e^{-\frac{(\enslabel-\enslabel_0)^2}{2\pwidth^2}}\;,\;\enslabel_0\in\mathbb{R},\;\pwidth\in\mathbb{R}^+ ,
\end{equation}
with the (dimensionless) mean $\lambda_0$ and the (dimensionless) width $\sigma$. The energy function  (\ref{eq:qubitphi}) and the decoherence rate  (\ref{eq:qubitgamma}) are computed from the characteristic function $\f_{\rm G}(\Deltaz t)= \exp[i\Deltaz\enslabel_0-\frac{1}{2}(\pwidth \Deltaz t)^2]$:
\begin{align}\label{eq:qubit-rates-gaussian}
 \varphi_{G} =\hbar \frac{\Deltaz}{2} \enslabel_0 \;\; ; \;\; \rate_{G}(t) = \frac{1}{2}(\Deltaz\pwidth)^2 t.
\end{align}
The energy function $\varphi_{G}$ is, as for the Cauchy-Lorentz distribution, time-independent and fixed by the average value $\enslabel_0$, which gives rise to a constant Larmor precession frequency $\Deltaz \enslabel_0$ (see Fig.~\ref{fig:qubit-all-coherent} of Appendix \ref{app:qubit-coherent}). This is a direct consequence of the symmetry of $p_{\rm G}(\enslabel)$ with respect to its mean. 
In addition, the Gaussian shape of the distribution brings about a positive, linearly in time increasing decoherence rate $\gamma_{\rm G} (t)$ whose slope scales quadratically with the variance $\pwidth$. We remark that such an above all bounds increasing decoherence rate would be considered unnatural from an open-system perspective, although it can be realized with suitable engineering of the environment \cite{Liu2011}; here, however, it arises as a natural and unavoidable consequence of the Gaussian ensemble average. On the level of the resulting average dynamics, the decoherence rate $\gamma_{\rm G}$ leads to a fast Gaussian decay of the coherences scaling with the variance squared, $\matelem{1}{\avgdm(t)}{2} \propto \exp\left[-1/2(\Deltaz \pwidth)^2 t^2 \right]$ (Fig.~\ref{fig:qubit-all}). Moreover, the positivity $\gamma_{\rm G} (t) \geq 0$ of the rate for all times $t\geq 0$ implies a strictly monotonous decay of the purity $\Tr{\avgdm^2(t)}$ \cite{Breuer2002}.

Furthermore, for the Gaussian distribution all cumulants of order $n>2$ vanish, $\kappa^{(n)}=0$. Hence, the short-time disorder master equation (\ref{eq:qubit short-time-diagonal-qme}) is exact for all times, because it captures the full dependence of the ensemble average dynamics upon the first two cumulants. From the Marcinkiewicz theorem \cite{Marcinkiewicz1939} it follows that the Gaussian distribution is the only possible distribution which has a finite number of non-zero cumulants, and thus it is the only distribution for which the short-time master equation (\ref{eq:qubit short-time-diagonal-qme}) remains exact at all times.

\subsection{Uniform box distribution}
The uniform box distribution is illustrative for continuous distributions with a finite support (cut-off). As we will see, such distributions with a cut-off naturally induce drastic dynamical consequences such as revivals, e.g., of coherences, as it becomes manifest by the time-dependence of the decoherence rates. The uniform box distribution $p_{\rm B}(\enslabel)$ is given by 
\begin{equation}\label{eq:qubit-distribution-box}
	p_{\rm B}(\enslabel)= 
	\left\{ \begin{array}{ll}
         \frac{1}{\pwidth} & \enslabel\in\left[\enslabel_0-\frac{\pwidth}{2},\enslabel_0+\frac{\pwidth}{2}\right]\\
        0 & \text{else} \end{array} \right. ;\enslabel_0\in\mathbb{R}, \pwidth\in\mathbb{R}^+ ,
\end{equation}
with the (dimensionless) location paramer $\lambda_0$ and the (dimensionless) scale parameter $\pwidth$. The corresponding characteristic function is 
$\f_{\rm B}(\Deltaz t) = \sinc \left((\pwidth \Deltaz t)/2\right) \exp[i \enslabel_0 \Deltaz t] $. For the energy function  (\ref{eq:qubitphi}) and the decoherence rate (\ref{eq:qubitgamma}) one thus obtains
\begin{align} \label{eq:single_qubit_uniform_box_decoherence_rate}
\varphi_{\rm B}= \hbar \frac{\Deltaz}{2} \enslabel_0 \;\; ; \;\; \rate_{\rm B}(t) =\frac{1}{2}\left(\frac{1}{t}-\frac{\pwidth\Deltaz}{2}\text{cot}\left(\frac{\pwidth \Deltaz}{2}  t\right)\right) .
\end{align}
As we can see, the energy function $\varphi_{\rm B}$ is once more time-independent and determined by the mean value $\enslabel_0$, which is due to the fact that the uniform box distribution is symmetric with respect to it. The coherent part of the dynamics is thus again a precession about the $z$-axis with constant Larmor frequency $\Deltaz \enslabel_0$. However, in contrast to the above examples of Cauchy-Lorentz and Gaussian distributions, 
the width $\pwidth$ now not only defines a dephasing rate, but also specifies the times of the singularities $\tau_n=2 n \pi/(\pwidth\Deltaz),\;n\in\mathbb{N}^+$, of the decoherence rate $\gamma_{\rm B}(t)$ arising from the box distribution (see Fig.\ref{fig:qubit-all}), the consequences of which shall be discussed below.

While the temporal behavior of the decoherence rate $\gamma_B(t)$ may be surprising, we emphasize that it is a direct consequence of the specific form of $p_{\rm B}$ and it particularly clearly reflects the underlying ensemble average dynamics. As is evident from Figure~\ref{fig:qubit-all}, whenever the decoherence rate diverges and jumps from positive to negative values, there is a transition from decaying to increasing coherences, i.e., revivals (see also Fig.~\ref{fig:qubit-all-coherent} of Appendix \ref{app:qubit-coherent}). At the times of divergence $\tau_n$, the inverse of the dynamical map $\Lambda_t^{-1}$ is not defined. This is because, at these times, the off-diagonal terms of the ensemble average state vanish exactly. Therefore, since the master equation is a first order differential equation, a diverging decoherence rate is required to induce a revival. The divergences are formally unproblematic, because the set of diverging points $\tau_n$ has measure zero and $\exp[-2\int_0^t \gamma_{\rm B}(t)] = \sinc \left((\pwidth \Deltaz t)/2\right)$ remains finite for all times $t\geq 0$. Thus, the solutions to the master equation remain bounded.

The coherence revivals described by $\gamma_{\rm B}(t)$ follow from the finite energy distribution width $\hbar\Deltaz\pwidth$; it has been known for long time that finite energy scales in quantum systems lead to recurrences \cite{Schrodinger1926,Bocchieri1957}. Thus, any disorder distribution with a finite or closed support gives rise to revivals. On top of the revivals, the finite energy width $\pwidth \hbar \Deltaz$ leads to a slow overall $\matelem{1}{\avgdm(t)}{2} \propto t^{-1}$ decay of the coherences.

Note that, despite the fact that $\gamma_{\rm B}(t)<0$ in some time intervals, the dynamics are physical at all times, by construction from the ensemble average. This is highlighted by the fact that $\int_0^t \gamma_{\rm B}(t')dt' > 0$ $\forall t> 0$, which can be shown to guarantee the complete positivity of the resulting dynamics using, e.g., the Jamilkowsky isomorphism \cite{Jamiolkowski1972}. The periodic divergences of the decoherence rate indicate that the dynamical map is not completely divisible. Both the negativity of the decoherence rate and the non-divisibility of the map reflect the occurrence of coherence revivals, a physical hallmark of non-Markovianity \cite{Breuer2015,Lambropoulos2000}.

We remark that the short-time master equation (\ref{eq:qubit short-time-diagonal-qme}) captures fully the coherent part of the average dynamics, as the exact energy function $\varphi_{\rm B}$ is time-independent. However, the decoherence rate $\gamma_{\rm B}(t)$ is approximated to first-order in time (cf.~Eq.~(\ref{eq:qubit_leading_order})). By direct comparison of the first and third order {(the second order vanishes) of the cumulant expansion (\ref{eq:qubitgamma_cumulant}) of $\gamma(t)$ in time, the range of validity of the short-time master equation is found to be restricted to times $t\ll \abs{2\dot{\gamma}_{\rm B}(t)/\dddot{\gamma}_{\rm B}(t)}^{1/2} \propto 1/(\Deltaz\pwidth)$. In other words, the validity range of the short-time master equation scales inversely proportional to the disorder strength. In particular, the short-time master-equation does not capture the revivals of the coherences at times $\tau_n \propto n/
(\Deltaz\pwidth)$, nor their subsequent algebraic decay.
 
 \subsection{L\'{e}vy distribution} \label{Sec:Levi}
The L\'evy distribution stands for distributions which may be asymmetric with respect to their median value, and which have slowly decaying tails. Such distributions have been proposed to describe the effective disorder mediated by slow changes in the structured background in molecular systems \cite{Barkai2000} and are generally abundant in complex systems \cite{Bardou2002}. The L\'{e}vy distribution $p_{\rm Le}(\enslabel)$ and its characteristic function $\f_{\rm Le}(\Deltaz t)$ are parametrized as
 \begin{equation}\label{eq:qubit-levy-distr}
	p_{\rm Le}(\enslabel)= 
	\sqrt{\frac{\pwidth}{2\pi}} \, \frac{e^{-\frac{\pwidth}{2(\enslabel-\enslabel_0)}}}{(\enslabel-\enslabel_0)^{3/2}}\;,\;\enslabel > \enslabel_0 \; , \; \pwidth\in\mathbb{R},\;\enslabel_0\in\mathbb{R}^+ ,
\end{equation}
 where $\sigma$ and $\lambda_0$ are the scale and the location parameter, respectively. This results in the characteristic function $\f_{\rm Le}(\Deltaz t) =\exp[i\enslabel_0 t-\sqrt{-2i\pwidth t}]$.  The energy function $\varphi_{\rm Le}(t)$ and the decoherence rate $\gamma_{\rm Le}(t)$ then follow as
\begin{align}\label{eq:qubit-levy-rates}
\varphi_{\rm Le}(t) = \hbar \frac{\Deltaz}{2}\enslabel_0 - \frac{\hbar}{4} \sqrt{\frac{\pwidth\Deltaz}{t}} \;\; ; \;\; \gamma_{\rm Le}(t) =\frac{1}{4} \sqrt{\frac{\pwidth\Deltaz}{t}} .
\end{align}
The location parameter $\enslabel_0$ once more defines the time-independent term of the coherent part of the master equation $\varphi_{\rm Le}$. On the other hand, the scale parameter, or more precisely its square-root $\sqrt{\pwidth}$, here not only fixes the rate of growth of the incoherent part $\gamma_{\rm Le}(t)$, but also sets a rate of growth of the time-dependent term of the coherent part $\varphi_{\rm Le}(t)$. This is, because, in contrast to the Cauchy-Lorentz, the Gaussian, and the uniform box distribution, the L\'{e}vy distribution is asymmetric, thus having non-vanishing odd cumulants. As a consequence, the energy function $\varphi_{\rm Le}(t)$ is time-dependent. The latter leads to an also time-dependent Larmor precession frequency (see Appendix \ref{app:qubit-coherent}). Figuratively speaking, thinking of spins $1/2$ in a magnetic field, the asymmetry of the L\'evy distribution (\ref{eq:qubit-levy-distr}) implies a statistical overweight of fast over slowly rotating spins, which induces on average a coherent rotation with a time-dependent frequency. The latter is mediated by the time-dependence of the energy function.
On top of the coherent dynamics, the L\'{e}vy distribution decoherence rate leads to an overall decay $\matelem{1}{\avgdm(t)}{2} \propto \exp\left[-\sqrt{t}\right]$, which is at small times faster than the Gaussian decay resulting from the Gaussian distribution, and at large times slower than the exponential decay resulting from the Cauchy-Lorentz distribution  (see Fig.~\ref{fig:qubit-all}). Since $\gamma_{\rm Le}(t)>0$ for all times $t$, the purity of the state $\Tr{\avgdm^2(t)}$ decays strictly monotonously.

Similarly to the Cauchy-Lorentz distribution, the L\'{e}vy distribution does not possess moments and cumulants. This explains why the short-time master equation (\ref{eq:qubit short-time-diagonal-qme}), predicting a time-independent energy function and a linearly-in-time increasing decoherence rate, does not capture correctly the $1/\sqrt{t}$ time-dependence of $\varphi_{\rm Le}(t)$ and $\gamma_{\rm Le}(t)$.
\begin{figure}[]
\includegraphics[width=0.95\columnwidth]{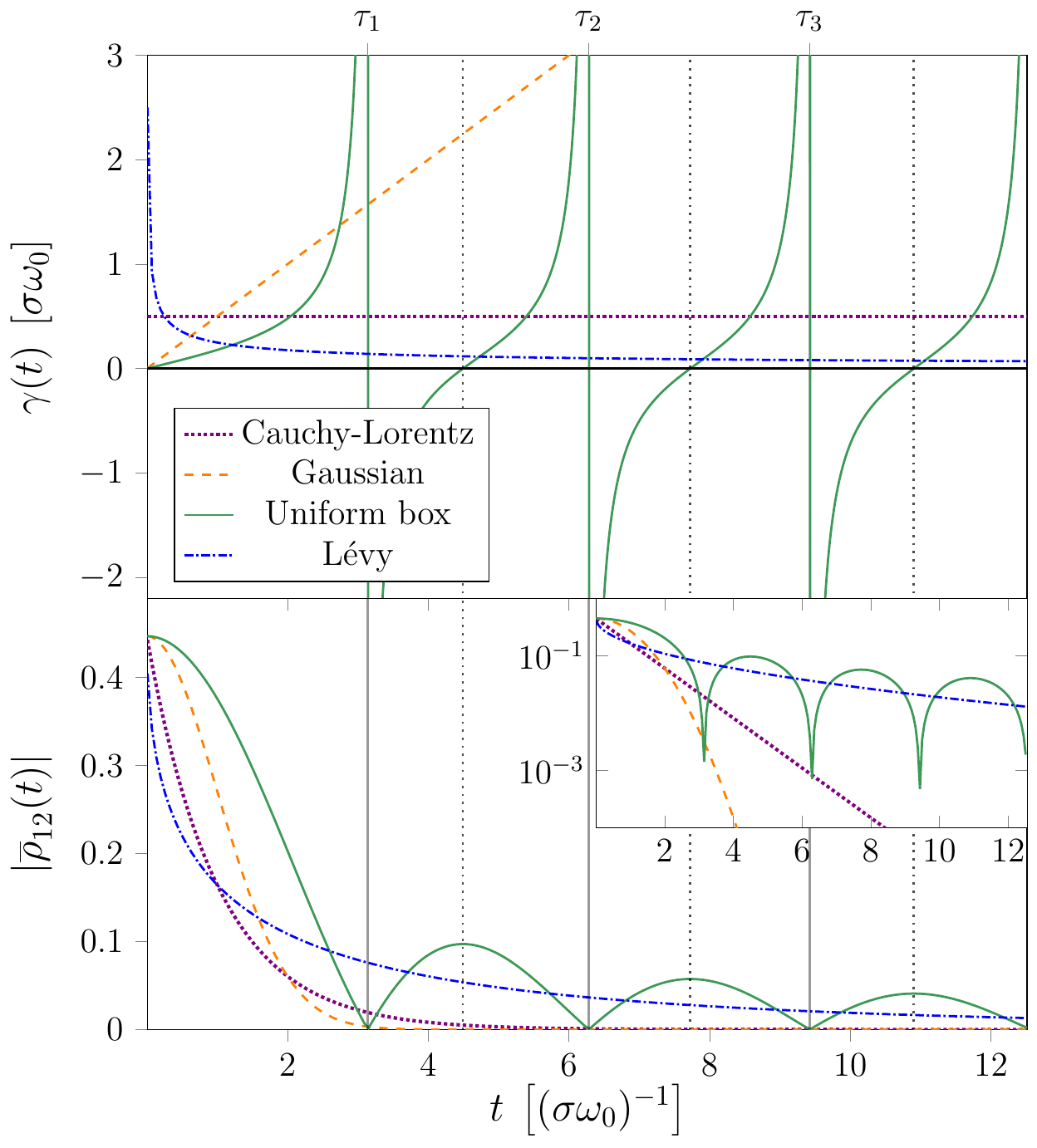}
	 \caption{[Color online] Ensemble average dynamics of a single qubit with spectral disorder. Top panel: Decoherence rates $\gamma_{\rm CL}(t),\gamma_{\rm G}(t),\gamma_{\rm Le}(t)$ for the Cauchy-Lorentz, the Gaussian, and the L\'{e}vy distribution with coinciding scale parameters $\pwidth$, and $\gamma_{\rm B}(t)$ for the uniform box distribution, for visual clarity with $2\pwidth$. Bottom panel: Decay of the coherences, i.e.,~of the off-diagonal terms of the average density matrix $\abs{\matelem{1}{\avgdm(t)}{2}}=\abs{\matelem{2}{\avgdm(t)}{1}}$, for the four distributions. The initial state is mixed and chosen so as to have no particular symmetry: $\dm(0) = (1/2) (\id+\vec{b}\vec{\sigma})$, with the initial Bloch vector $\vec{b} = (2/5,4/5,1/3)$, $\vec{\sigma}\define (\sigma_x,\sigma_y,\sigma_z)$ the vector of Pauli matrices, and the resulting initial coherence $\abs{\matelem{1}{\avgdm (0)}{2}} \approx 0.43$. All four distributions lead to dephasing dynamics, as reflected by the Lindblad operators of the disorder master equation (\ref{eq:qubitqme}). The coherence revivals in case of the uniform box distribution (green solid line) are manifestly in correspondence with the divergences, at times $\tau_{i}$ (full, vertical, gray lines), and zeros (dotted, vertical, gray lines) of the decoherence rate $\gamma_{\rm B}(t)$. Inset: Semi-logarithmic plot showing the decay of the coherence's envelopes as $\exp\left(-t\right)$, $\exp\left(-t^2\right)$, $1/t$ and $\exp\left(-\sqrt{t}\right)$, respectively.
	 }
	 \label{fig:qubit-all}
\end{figure}


\section{Spectral disorder in $d$ dimensions}\label{sec:diaognal-general-disorder}
We now consider the general, $d$-dimensional case of spectral disorder (with identical eigenvectors for all realizations, see Sec.~V above). Therewith, it is for example possible to describe higher-dimensional spin states, 
or composite systems such as an ensemble of two-level atoms \cite{Carnio2015}. As we shall see, the particular structure of the spectral disorder directly translates into the form of the Lindblad operators. 

The disorder ensemble consists of $d \times d$ Hamiltonians which all mutually commute, to guarantee the existence of 
a common eigenbasis $\{\ket{j}\}_{j=1}^d$ (cf.~Eq.~(\ref{eq:diagonal-H})),
\begin{align} \label{eq:diagonal-H-2}
	\ens{\left(  \H_{\vec{\enslabel}} = \hbar \Deltaz \sum_{j=1}^d \enslabel_j  \ketbra{j}{j} \; , \; p_{\vec{\enslabel}} = p(\enslabel_1,...,\enslabel_d)\right)}.
\end{align}
 The disorder is then fully characterized by the eigenvalue distribution $p_{\vec{\enslabel}} = p(\enslabel_1,...,\enslabel_d)$  ($\vec{\enslabel}\define \left(\enslabel_1,...,\enslabel_d\right))$. Analogously to the single-qubit case, in the common eigenbasis $\ens{\ket{j}}_{j=1}^d$ the diagonal terms of the ensemble average density matrix $\avgdm (t)$ are time-independent, whereas the off-diagonal terms evolve according to $\matelem{j}{\overline{\dm}(t)}{k} = \matelem{j}{\overline{\exp\left[-i (\enslabel_j-\enslabel_k)\Deltaz t\right]}}{k}$, $k \ne j$. Hence, the average dynamical matrix $\dynmat(t)$, Eq.~(\ref{eq:dynamical matrix}), is given by
\begin{align}\label{eq:F d-diag}
	\dynmat_{jk,rs}(t) = \delta_{jr}\delta_{ks} \f^*_{jk}(\Deltaz t),
\end{align}
with the characteristic function of the level-spacing distribution $\tilde{p}_{j k}(\Delta \enslabel) = \int_{-\infty}^{\infty} d\enslabel_j  \int_{-\infty}^{\infty}  d\enslabel_k \, p(\enslabel_1,...,\enslabel_d) \delta(\enslabel_j-\enslabel_k - \Delta \enslabel)$ defined by
\begin{align} \label{eq:diagonal-characteristic-function}
	\f_{jk}(\Deltaz t) &\define \int_{-\infty}^{\infty} d\Delta \enslabel  \, \tilde{p}_{j k}(\Delta \enslabel) e^{i \Delta \enslabel \Deltaz t} \\
	&= \int_{-\infty}^{\infty} d\enslabel_j  \int_{-\infty}^{\infty}  d\enslabel_k \, p(\enslabel_1,...,\enslabel_d) e^{i (\enslabel_j-\enslabel_k) \Deltaz t}. \nonumber
\end{align}
The representational matrix $\qmemat(t)$ of the master equation, Eq.~(\ref{eq:matrixqme}), consequently reads
\begin{align}\label{eq:diagonal-G(t)}
	\qmemat_{jk,rs}(t) =\sum_{m,n=1}^d \dot{\dynmat}_{jk,mn}(t) \dynmat^{-1}_{mn,rs}(t) = \delta_{jr}\delta_{ks}\frac{\dot{\f}^*_{jk}(\Deltaz t)}{\f^*_{jk}(\Deltaz t)} \, ,
\end{align}
and the non-Lindblad form of the disorder master equation, Eq.~(\ref{eq:qmewithG}), is given by
\begin{align} \label{eq:diagonal-projector-qme}
	 	\dot{\avgdm}&= \sum_{j,k=1}^d \left[ \frac{d}{dt}\ln(\f^*_{jk}(\Deltaz t)) \right] \Pi_{j}\avgdm \Pi_{k}^\dagger ,
\end{align}
with $\Pi_j\define \ketbra{j}{j}$ the projectors onto the common eigenvectors of the Hamiltonians $H_\enslabel$.

Before deriving the final Lindblad form, it is convenient to compare the short-time expansion of  (\ref{eq:diagonal-projector-qme}) with the short-time disorder master equation (\ref{eq:short-time_master_equation}). The leading-order in time approximation of $(d/dt)\ln(\f^*_{jk}(t))$ in (\ref{eq:diagonal-projector-qme}) reads
\begin{align} \label{eq:short-time-diag_rate}
\frac{d}{dt}\ln(\f^*_{jk}&(\Deltaz t)) = -i\Deltaz\left(\avg{\enslabel}_j-\avg{\enslabel}_k\right) \\
	& + t\Deltaz^2\left((\avg{\enslabel}_j-\avg{\enslabel}_k)^2-\avg{(\enslabel_j-\enslabel_k)^2}\right) + \mathcal{O}(t^2) . \nonumber
\end{align}
Inserting (\ref{eq:short-time-diag_rate}) into (\ref{eq:diagonal-projector-qme}), one can deduce the non-diagonal Lindblad form for the short-time approximation:
\begin{align} \label{eq:short-time-diagonal-qme}
	\dot{\avgdm}& = -\frac{i}{\hbar} \left[\avg{\H},\avgdm\right] \\
	&+2\Deltaz^2 t \sum_{j,k=1}^d (\avg{\enslabel_j\enslabel_k}-\avg{\enslabel}_j\avg{\enslabel}_k) \Bigl(\Pi_j\avgdm \Pi_k-\frac{1}{2}\left\{\Pi_j\Pi_k,\avgdm\right\}\Bigr) \nonumber.
\end{align}
On the other hand, with the ensemble-averaged Hamiltonian $\avg{\H} = \hbar\Deltaz \sum_{j=1}^d \avg{\enslabel}_j\ketbra{j}{j}$, one obtains for the short-time Lindblad operators and decay rates (\ref{eq:short-time_Lindbladians})
\begin{align} \label{eq:diagonal-short-time-operators}
	L_{\vec{\enslabel}} = \sum_{j=1}^d (\enslabel_j-\avg{\enslabel}_j) \ketbra{j}{j} \;\; ; \;\;
	\gamma_{\vec{\enslabel}}(t) = 2p_{\vec{\enslabel}}\;\Deltaz^2 t.
\end{align}
Inserting (\ref{eq:diagonal-short-time-operators}) into (\ref{eq:short-time_master_equation}) and performing the integrals over the disordered eigenvalues, we find that, as expected, the short-time master equation coincides with the leading-order expansion (\ref{eq:short-time-diagonal-qme}) of the master equation.

We proceed now towards the general Lindblad form of the spectral disorder master equation. Following the matrix approach outlined in Section \ref{sec:methodology}, we obtain in terms of a Hermitian, traceless operator basis $\ens{\A_j}_{j=0}^{d^2-1}$ the effective Hamiltonian $\H(t) = (i\hbar/2) \left( \Cred(t)-\Cred^\dagger(t) \right)$ (cf.~Eq.~(\ref{eq:C-Hamiltonian})) with
\begin{align} \label{eq:diagonal-Ctilde}
	\Cred(t) =\sum_{j,k=1}^d  \left[ \frac{d}{dt}\ln(\f^*_{jk}(\Deltaz t))\right]\Bigl(\frac{\id}{2d^2}+\frac{1}{\sqrt{d}}\sum_{m=1}^{d^2-1}\A_m\A_{m,jj}\Bigr),
\end{align}
and the $(d^2-1)\times(d^2-1)$ decoherence matrix (see Eq.~(\ref{eq:C matrix}))
\begin{align} \label{eq:diagonal-decoherence-matrix}
	\Gamma_{mn}(t) \define \sum_{j,k=1}^{d}  \left[\frac{d}{dt}\ln(\f^*_{jk}(\Deltaz t)) \right] \A_{m,jj}\A_{n,kk}.
\end{align}
Note that in Eq.~(\ref{eq:diagonal-Ctilde}) and in Eq.~(\ref{eq:diagonal-decoherence-matrix}) only the diagonal terms of the operators $\A_m$ are relevant. Henceforth, a suitable operator basis is given by the Gell-Mann matrices, cf.~Eq.~(\ref{eq:Gell-Mann-matrices}), since only the $d-1$ diagonal operators $\gm_d^l$ have non-vanishing diagonal elements in the state basis defined by the disorder ensemble, Eq.(\ref{eq:diagonal-H-2}). More precisely, we get
\begin{equation}\label{eq:diagonal-amjj}
	\A_{m,jj} = \left\{ \begin{array}{ll}
        \matelem{j}{\A_m}{j} & \A_m \in \left\{\gm_d\right\}\\
        0 & \text{else} \end{array} \right. .
\end{equation}
Since $\phi_{jk}=\phi_{kj}^*$ and $\A_{m,jj}\in\mathbb{R}$, the decoherence matrix $\Gamma(t)=\Gamma^\dagger(t)$  is, as expected, Hermitian and can therefore be diagonalized for given characteristic functions $\phi_{jk}(\omega_0 t)$. The resulting Lindblad operators in the diagonal form of the master equation are then linear combinations of the $\gm_d^l$ and thus also diagonal operators. Consequently, general spectral disorder always results in dephasing dynamics in the state basis $\ens{\ket{j}}_{j=1}^d$ defined by the disorder ensemble. It is thus a natural generalization of the single qubit with spectral disorder considered in the previous section. The populations are time-independent, whereas the off-diagonal terms of the density matrix evolve non-unitarily. This result can be understood as follows: For each single realization of the disorder, the populations of the individual Hamiltonians' eigenstates are time-independent and thus also remain time-independent after the averaging, whereas the oscillating off-diagonal terms undergo dephasing due to the averaging over the disorder in the eigenvalues.

Although there is no direct cross-talk between the coherences on the level of the master equation, their dynamics may exhibit strong correlations, mediated by the correlations among the elements of the decoherence matrix. Furthermore, since the decoherence matrix depends only on the level-spacing distribution, all disorder distributions $p_{\vec{\enslabel}}$ which give rise to the same level spacing statistics $\tilde{p}_{j k}(\Delta \enslabel)$ induce the same disorder master equation.

In the following, we have a more detailed look at two specific types of spectral disorder, namely global spectral disorder, which is characterized by a single disorder parameter, and fully uncorrelated spectral disorder.

\subsection{Global spectral disorder}

Global spectral disorder represents a direct generalization of the previously studied spectrally disordered single qubit to $d$ dimensions, in the sense that the spectrum is only subject to a single disorder parameter. This may for example describe an array of $N$ uncoupled two-level atoms in a common, static magnetic field, which varies slowly in intensity over time, giving rise to collective dephasing \cite{Carnio2015}.

We consider the ensemble of random Hamiltonians
\begin{align}\label{eq:diagonal-global-ensemble}
		\ens{\left( \H_\enslabel = \enslabel \H_0 = \enslabel \sum_{j=1}^d \hbar \omega^0_{j} \ketbra{j}{j} \; , \; p_\enslabel=p(\enslabel)\right)} ,
\end{align}
with $\ens{\ket{j}}_{j=1}^d$ the common eigenbasis and $\hbar \omega^0_j$ the eigenvalues of the reference Hamiltonian $\H_0 = \sum_{j=1}^d \hbar \omega^0_{j} \ketbra{j}{j}$. The single disorder parameter $\enslabel$ is distributed according to $p(\enslabel)$. Note that, for convenience, the global disorder definition (\ref{eq:diagonal-global-ensemble}) differs from the definition (\ref{eq:diagonal-H-2}) for general spectral disorder. The two definitions correspond via the identification $\Deltaz \enslabel_j= \omega^0_j \enslabel$, which relates the probability distributions via $p_{\vec{\enslabel}} = \{ p(\enslabel) \; \text{if} \; \enslabel_j = \enslabel (\omega^0_j/\Deltaz); 0 \; \text{otherwise} \}$. In terms of the more natural characterization (\ref{eq:diagonal-global-ensemble}) of global spectral disorder, the dynamics are characterized by a single characteristic function $\f$ evaluated at the reference Larmor frequency differences $\omega^0_{jk}\define \omega_j^0-\omega_k^0$, which now mediate the dependence upon the indices $j,k$, 
\begin{align}\label{eq:diagonal-global-characteristic-function}
	\f(\omega^0_{jk}t) = \int_{-\infty}^\infty d\enslabel \, p(\enslabel) e^{i\enslabel\omega^0_{jk}t} .
\end{align}
All previous results for general spectral disorder can now be taken over
by identifying the characteristic functions $\f_{jk}(\Deltaz t)$, (\ref{eq:diagonal-characteristic-function}), with $\f(\omega^0_{jk}t)$, (\ref{eq:diagonal-global-characteristic-function}).

As a consequence of the single joint disorder parameter for all eigenvalues, the dynamics of the coherences are strongly correlated in terms of the characteristic function (\ref{eq:diagonal-global-characteristic-function}), $\matelem{j}{\avgdm(t)}{k} \propto\f^*(\omega^0_{jk} t)$. On the other hand, they also depend on the Larmor frequency difference $ \omega^0_{jk} $ connecting the levels $j$ and $k$, which renormalizes their decoherence time. Therefore, all coherences will decay with the same envelop function, but at different velocities.

Interestingly, if there are degeneracies in the spectrum of the reference Hamiltonian $\H_0$, i.e.,~at least two eigenvalues $\hbar\omega_j^0$ and $\hbar\omega_k^0$ coincide, the off-diagonal terms of the density matrix $\avgdm(t)$ associated with these degenerate eigenvalues are completely time-independent ($\dot{\f}(\omega^0_{jk}t)=0$) and, hence, survive the averaging procedure, even in the asymptotic limit. This mechanism can for example be exploited in order to find long-lived entangled states \cite{Carnio2015}.

Finally, it is obivous that the characteristic function (\ref{eq:diagonal-global-characteristic-function}) is the same as the characteristic function (\ref{eq:qubitf(t)}) for a single qubit with spectral disorder, given the probability distributions coincide. Thereby, the Hamiltonian $\H(t)$ and the decoherence matrix $\Gamma_{mn}(t)$ are for global spectral disorder directly obtained by identifying the real and imaginary part of $-\frac{d}{dt}\ln [\f^*(\omega^0_{jk} t)]$ in Eq.~(\ref{eq:diagonal-Ctilde}) and Eq.~(\ref{eq:diagonal-decoherence-matrix}) with the qubit decoherence rate $2\gamma_{jk}(t)$, Eq.~(\ref{eq:qubitgamma}) and the qubit energy function $(2/\hbar)\varphi_{jk}(t)$, Eq.~(\ref{eq:qubitphi}), respectively, where the indices $j,k$ indicate that $\Deltaz $ is replaced by $\omega^0_{jk}$.

In particular, for a Gaussian distribution $p_G(\enslabel)$ with width $\sigma$, cf.~Eq.~(\ref{eq:qubit-distribution-gaussian}), the qubit rate and energy function (\ref{eq:qubit-rates-gaussian}) yield $-2\frac{i}{\hbar}\varphi_{jk}(t)-2\gamma_{jk}(t) =- i\avg{\enslabel} \omega^0_{jk} - (\sigma {\omega^0_{jk}})^2 t$. One can then easily derive a diagonal Lindblad master equation 
\begin{align} \label{eq:diagonal-global-gaussian-qme}
	\dot{\avgdm}=-\frac{i}{\hbar} \avg{\enslabel} \left[\H_0,\avgdm\right] +2\frac{\sigma^2}{\hbar^2}t \Bigl(\H_0\avgdm\H_0^\dagger - \frac{1}{2}\left\{\H_0^\dagger\H_0,\avgdm\right\}\Bigr),
\end{align}
where the single Lindbald operator is given by the reference Hamiltonian $H_0$. Since the global spectral disorder master equation for a Gaussian distribution, Eq.~(\ref{eq:diagonal-global-gaussian-qme}), is at most linear in time, it coincides with the corresponding short-time master equation obtained from (\ref{eq:short-time-diagonal-qme}). We thus find that in this case the short-time dynamics remains valid at arbitrary times. Interestingly, due to the absence of higher cumulants, the master equation can be generalized to time-dependent noise $\enslabel(t)$, as was studied in \cite{Benatti2012}.

\subsection{Uncorrelated spectral disorder}

We finally consider a spectral disorder ensemble (\ref{eq:diagonal-H-2}) with a fully uncorrelated eigenvalue distribution,
\begin{align}
	p_{\vec{\enslabel}}=p_1(\enslabel_1)...p_d(\enslabel_d),
\end{align}
 where the probability distributions $p_j$, $j = 1,\dots,d$, may in general differ from one another. The characteristic function (\ref{eq:diagonal-characteristic-function}) then factorizes and $(d/dt)\ln( \f^*_{jk}(\omega_0 t)) = (d/dt) \left[\ln\left(\f^*_j(\omega_0 t)\right) +\ln\left(\f_k(\omega_0 t)\right) \right]$
with $\f_j(\Deltaz t) = \int_{-\infty}^\infty d\enslabel_j p_j(\enslabel_j) \text{exp}\left[i \enslabel_j\Deltaz t\right]$.
The characteristic function $\f_j(\Deltaz t)$ specifies the Hamiltonian via (\ref{eq:diagonal-Ctilde}) and the decoherence matrix via (\ref{eq:diagonal-decoherence-matrix}). We can then derive a diagonal Lindblad form of the master equation,
\begin{align}\label{eq:diagonal-uncorrelated-qme}
	\dot{\avgdm}= -\frac{i}{\hbar}\left[\H_{\rm}(t),\avgdm\right]+\sum_{j=1}^d \gamma_{j}(t) \left(\Pi_j\avgdm\Pi_j -\frac{1}{2}\left\{\Pi_j\Pi_j,\avgdm\right\}\right) ,
\end{align}
where $\H_{\rm}(t) = -\hbar\sum_{j=1}^d\Im\left[\frac{d}{dt}\ln\left(\f^*_j(\omega_0 t)\right)\right]\Pi_j $ and the $\Pi_j=\ketbra{j}{j}$ are again the projectors onto the common eigenvectors of the Hamiltonians in the ensemble.
Since there are no correlations in the considered spectral disorder, each decoherence rate is equal to $\gamma_j(t)\define -2\Re\left[\frac{d}{dt}\ln\left(\f^*_j(\omega_0t)\right)\right]$ and thus depends only on the characteristic function corresponding to the $j^{\text{th}}$ eigenvalue. Hence, as opposed to the previous strongly correlated case of global spectral disorder, the different coherences will in general exhibit disparate decay patterns. Note that the above diagonal Lindblad form (\ref{eq:diagonal-uncorrelated-qme}) can also be directly derived from the projector form (\ref{eq:diagonal-projector-qme}).


\section{Unitarily invariant disorder}\label{sec:random matrics}
In the previous section, we studied the ensemble average dynamics resulting from spectral disorder, where all Hamiltonians in the ensemble shared a common eigenbasis. We now generalize to the case that also the eigenstates are subject to disorder. For example, think again of the ensemble of spins $1/2$ in a static magnetic field from Section \ref{sec:diagdisorder}, but now in addition to the magnitude also the orientation of the field is random. The latter can for instance be induced by local impurities. More generally speaking, disorder in the eigenstates is implied whenever the Hamiltonians in the ensemble (\ref{eq:disorder_ensemble}) do not all commute with one another. This is, for instance, typical for systems with a kinetic hopping term and a disordered on-site potential, such as, e.g., the Anderson model \cite{Anderson1958}.
 In the following we elaborate how disorder, not only in the eigenvalues, but also in the eigenvectors, can affect the structure of the master equation describing the ensemble average. Concretely, we focus on an isotropically randomized eigenstate distribution.

Let us consider an ensemble of $d\times d$ Hamiltonians with uniformly distributed eigenvectors and with eigenvalues 
distributed independently of the eigenvectors,
\begin{align} \label{eq:unitarily_invariant_disorder_ensemble}
	\ens{\left( \H_{W, \vec{\enslabel}} = W D_{\vec{\enslabel}} W^{\dagger} \; , \; p_{W, \vec{\enslabel}} = p_{\vec{\enslabel}} \right)},
\end{align}
where $D_{\vec{\enslabel}} = \hbar \Deltaz \text{diag}(\enslabel_1,...,\enslabel_d)$ and $W W^\dagger = \id$. The isotropic eigenvector distribution is reflected by the invariance of the probability distribution $p_{\vec{\enslabel}}$ with respect to the unitary transformations $W$, $p_{\id,\vec{\enslabel}} = p_{W,\vec{\enslabel}} = p_{\vec{\enslabel}}$. Prominent members of the disorder type (\ref{eq:unitarily_invariant_disorder_ensemble}) are the unitarily invariant random matrix ensembles \cite{Haake2010,Stockmann2007,Mehta2004}. Generally speaking, random matrices have a wide range of application in the study of the statistical properties of complex quantum systems \cite{Guhr1998,Dyson1972, Walschaers2015,Gorin2008}. 

In order to understand the impact of the uniform eigenvector distribution on the ensemble average dynamics, we now derive the  master equation for the unitarily invariant disorder ensemble. The evolution of a single realization is given by $\dm_{W,\vec{\enslabel}}(t)= W \exp\left[-(i/\hbar) D_{\vec{\enslabel}} t\right] W^\dagger \dm(0) W \exp\left[(i/\hbar)D_{\vec{\enslabel}} t\right] W^\dagger$. The ensemble average state is then obtained by integrating both over the unitaries $W$ in terms of the Haar measure $d_\mu (W)$, and over the eigenvalue distribution $p_{\vec{\enslabel}} = p(\enslabel_1,...,\enslabel_d)$:
\begin{align} \label{eq:rnd mat tmap}
	\avgdm(t) &= \int d \vec{\enslabel} \: p_{\vec{\enslabel}}  \int d_\mu (W)  W e^{-\frac{i}{\hbar} D_{\vec{\enslabel}} t} W^\dagger \dm(0) W e^{\frac{i}{\hbar} D_{\vec{\enslabel}} t} W^\dagger .
\end{align}
The Haar measure integral in Eq.~(\ref{eq:rnd mat tmap}) can be conducted using the two following results which are obtained using the Weingarten calculus for unitary groups \cite{Gessner2013,Collins2006},
\begin{align}\label{eq:rnd-E2[x]}
	\int_{-\infty}^{\infty} d_\mu(W) W X W^\dagger = \frac{\Tr{X}}{d} \cdot\id 
\end{align}
and
\begin{align} \label{Eq:rm_integral}
	\int_{-\infty}^{\infty} d_\mu(W) &W X_1 W^\dagger X_2 W  X_3 W^\dagger= \nonumber\\ & \frac{d\Tr{X_1X_3}-\Tr{X_1}\Tr{X_3}}{d(d^2-1)} \Tr{X_2} \cdot \id \nonumber\\&+ \frac{d\Tr{X_1}\Tr{X_3}-\Tr{X_1X_3}}{d(d^2-1)} \cdot X_2. 
\end{align}
Using these, one obtains for the time evolution of the ensemble-average state the dynamical map
\begin{align}\label{eq:rnd-depolarization-channel}
	\avgdm(t) &= \tmap_t\left[\dm(0)\right] = (1-a(t))\frac{\id}{d} + a(t)\dm(0) .
\end{align}
It describes a depolarization channel, i.e.,~the mixing of the initial state $\dm (0)$ with the maximally mixed state $\id/d$, with the time-dependent mixing probability $a(t)$ ($a(0)=1$); the latter is determined by the eigenvalue distribution $p_{\vec{\enslabel}}$ in terms of the sum over the characteristic functions for all level spacings $(\enslabel_j-\enslabel_k)$, Eq.~(\ref{eq:diagonal-characteristic-function}), which we denote as $\frnd(t) \define \frac{1}{d^2} \sum_{j,k=1}^d \phi^*_{jk}(\Deltaz t)$,
\begin{align} \label{eq:rndmat a(t)}
a(t) &\define  \frac{d^2 \frnd(t)-1}{d^2-1} .
\end{align}
Note that here the bar indicates averaging with respect to the eigenvalue distribution $p_{\vec{\enslabel}}$ only. Following the general procedure outlined in Section~\ref{sec:methodology}, one obtains the non-Lindblad master equation for unitarily invariant disorder,
\begin{align}\label{eq:rndmat qme}
	\dot{\avgdm} = \frac{\dot{a}(t)}{a(t)} \left(\avgdm-\frac{\id}{d}\right) . 
\end{align}
The non-Lindblad master equation (\ref{eq:rndmat qme}) is described by the representation matrix $\qmemat_{jkrs} = (\dot{a}(t)/a(t))(\delta_{jr}\delta_{ks} -(1/d)\delta_{jk}\delta_{rs})$, cf.~Eq.~(\ref{eq:qmewithG}). In Section \ref{sec:methodology}, we provided a systematic method to derive the general diagonal Lindblad form based on the representation matrix $\qmemat$. In the present case, however, it is possible to deduce the diagonal Lindblad form
\begin{align} \label{eq:rnd-lindblad-form-qme}
	\dot{\avgdm} &= \gamma(t) \sum_{j} \Big( L_{j}\avgdm L_{j}^\dagger -\frac{1}{2}\left\{L_{j}^\dagger L_{j},\avgdm\right\} \Big)
\end{align}
directly from Eq.~(\ref{eq:rndmat qme}) by requiring that the Lindblad operators satisfy $\sum_j L_{j}(t) \avgdm(t) L_{j}^\dagger = \id/d$ and $\sum_j L_{j}^\dagger L_{j} = \id$ (i.e.~the Lindblad operators form a basis in operator space). The single depolarization rate $\gamma(t)$ is then given by
\begin{align}\label{eq:rnd rate}
	\gamma(t) &= -\frac{d}{dt}\ln\left(a(t)\right)  = \frac{d^2\frnd(t)}{1-d^2\frnd(t)}. 
\end{align}
A possible choice for the Lindblad operators $L_j$ are the Hermitian Gell-Mann matrices $\ens{\gm^j}_{j=1}^{d^2-1}$ \cite{Bertlmann2008} (see also Eq.(\ref{eq:Gell-Mann-matrices})), complemented by the identity matrix and with a suitable normalization
\begin{align}
	\Lop_{0} = \frac{1}{d} \id \;\; ; \;\;  \Lop_{j} = \frac{1}{\sqrt{2d}}  \gm^j \; ; \; j = 1,...,(d^2-1).  
\end{align}
We thus find that the master equation (\ref{eq:rnd-lindblad-form-qme}) for unitarily invariant disorder does, first of all, not have a coherent Hamiltonian term. This is because averaging over a uniform distribution of eigenvectors cancels out all the terms arising from the (a-)symmetry in the eigenvalue distribution, i.e., the odd moments and cumulants, that generate the coherent dynamics (see Section \ref{sec:diagdisorder}). Moreover, the Lindblad operators $L_j$ are independent of the eigenvalue distribution; in other words, they are completely specified by the uniform eigenvector distribution alone, describing the isotropic dynamics towards the maximally mixed state which is characteristic of depolarization. The single depolarization rate $\gamma(t)$, on the other hand, is characterized by the eigenvalue distribution $p_{\vec{\enslabel}}$, more specifically by the level-spacing distribution. Therefore, different eigenvalue spacing statistics will lead to different mixing probabilities, as we will show in the examples below.

Before, it is interesting to see how the structure of the master equation (\ref{eq:rnd-lindblad-form-qme}) is recovered in the short-time limit from the general expression (\ref{eq:short-time_master_equation}). In the presently considered case of a unitarily invariant ensemble (\ref{eq:unitarily_invariant_disorder_ensemble}), one obtains, using (\ref{eq:rnd-E2[x]}), that the average Hamiltonian is $\avg{H} = \int d{\vec{\enslabel}}\int d_\mu(W) \H_{W,\vec{\enslabel}} = \Tr{\avg{D}_{\vec{\enslabel}}} \id/d$. According to (\ref{eq:short-time_Lindbladians}) the short-time decay rate and the Lindblad operators are thus given by
\begin{align}\label{eq:rnd-short-time-Lindbladian}
L_\enslabel = \frac{1}{\hbar\Deltaz} \Bigl( H_{W,\vec{\enslabel}} - \Tr{\avg{D}_{\vec{\enslabel}}} \frac{\id}{d} \Bigr) \, ; \, \gamma_{\vec{\enslabel}}(t) = \frac{2 \Deltaz^2 t}{\hbar^2} p_{\vec{\enslabel}} \, .
\end{align}
With (\ref{eq:rnd-short-time-Lindbladian}), and using $\int d_\mu(W) W X_1 W^\dagger X_2 = (\Tr{X_1}/d)X_2$ and (\ref{Eq:rm_integral}), one can perform the Haar measure integral in Eq.~(\ref{eq:short-time_master_equation}). Based on a suitable basis change, we then obtain the short-time diagonal Lindblad master equation
\begin{align}\label{eq:rnd-short-time-qme}
	\dot{\avgdm} = \frac{2 t \Deltaz^2}{(d^2-1)} &\left(\sum_{j,k=1}^d\avg{\enslabel_j\enslabel_k} -d\sum_{j=1}^d \avg{\enslabel_j^2}\right) \\
	&\times \sum_{j} \Big( L_{j}\avgdm L_{j}^\dagger -\frac{1}{2}\left\{L_{j}^\dagger L_{j},\avgdm\right\} \Big). \nonumber
\end{align}
This short-time master equation, which is obtained with the method outlined in Section \ref{sec:short-time_master_equation}, is in agreement with the leading-order expansion in time of the master equation (\ref{eq:rnd-lindblad-form-qme}) for $t \ll 1/\Deltaz$.

Note that the Lindblad term of the short-time master equation (\ref{eq:rnd-short-time-qme}) is composed of the same Lindblad operators $L_j$, which are determined only by the uniform distribution of the eigenvectors, as the finite-time master equation (\ref{eq:rnd-lindblad-form-qme}). On the other hand, the decoherence rate of the short-time description in (\ref{eq:rnd-short-time-qme}) contains only the first-order in time term of the exact rate (\ref{eq:rnd rate}). Therefore, differences between both equations must stem from the eigenvalue distribution specifying $\gamma(t)$. Possible deviations of the short-time approximation from the exact dynamics at larger times can thus be traced back to $p_{\vec{\enslabel}}$. By direct comparison of the first-order and third-order (second order vanishes) terms of the short-times expansion of $\gamma(t)$, we derive that the validity of the short-time master equation is limited to times 
\begin{align}\label{eq:rnd-short-time-validity}
 t \ll \abs{2\dot{\gamma}(t)/\dddot{\gamma}(t)}^{1/2} \propto \abs{\frac{\sum_{j,k=1}^d \avg{(\enslabel_j-\enslabel_k)^2}}{\Deltaz^2\sum_{j,k=1}^d \avg{(\enslabel_j-\enslabel_k)^4}}}^{1/2}, 
 \end{align}
which, as we will see in the example below, may be shorter than $1/\Deltaz$.

In summary, for unitarily invariant disorder, the distribution of the eigenvectors fully characterizes the Lindblad operators, i.e., the character of the decoherence process, whereas the distribution of the eigenvalues determines the temporal evolution of the corresponding decoherence rates. We will now treat in more detail the examples of the Gaussian unitary ensemble (GUE), which is often employed to uncover generic features of quantum chaotic systems \cite{Bohigas1984}, and, for comparison, the "Poissonian ensemble" (PE), which is meant to model an integrable counterpart \cite{Haake2010}. These ensembles are distinguished in their distribution of the eigenvalues $p_{\vec{\enslabel}}$ and, as we shall see, the corresponding master equations are accordingly distinguished by the time-dependence of their decoherence rates.

\subsection{Poissonian ensemble}

As a first example, we consider an uncorrelated eigenvalue distribution with each eigenvalue $\hbar \Deltaz \enslabel_j$ following a uniform box distribution $p_B(\enslabel_j)$ of width $\pwidth$ (cf.~Eq.~(\ref{eq:qubit-distribution-box})),
\begin{align}\label{eq:rnd-poisson-distribution}
	p_{\vec{\enslabel}}= \Pi_{j=1}^d P_{\rm B} (\enslabel_j).
\end{align}
This gives rise to the Poisson distributed eigenlevel spacing of the PE. To determine the depolarization rate (\ref{eq:rnd rate}), we note that for permutation symmetric distributions such as (\ref{eq:rnd-poisson-distribution}) one has
\begin{align} \label{eq:rndmat computef}
\frnd(t) = \frac{1}{d} + \frac{1}{d^2} \int d\enslabel_1 \int d\enslabel_2 e^{-i\Deltaz(\enslabel_1-\enslabel_2)t} R_2(\enslabel_1,\enslabel_{2}) ,
\end{align}
where
\begin{align}\label{eq:rnd 2point correlations}
R_2(\enslabel_1,\enslabel_2) = \sum_{i,j\,(i\neq j)} \avg{\delta\left(\enslabel_1-\enslabel_i\right) \delta\left(\enslabel_2-\enslabel_j\right)}
\end{align}
is the two-point correlation function of the distribution $p_{\vec{\enslabel}}$  \cite{Mehta2004}. For the Poissonian ensemble with the probability distribution (\ref{eq:rnd-poisson-distribution}), the two-point correlation function can be evaluated analytically and reads $R_2(\enslabel_1,\enslabel_2) = d(d-1)p_{\rm B}(\enslabel_1)p_{\rm B}(\enslabel_2)$. One then obtains the depolarization rate (\ref{eq:rnd rate}) 
\begin{align} \label{eq:rnd-poisson-rate}
	\gamma_{\rm P}(t) &=
	 -\frac{d (\frac{\Deltaz \pwidth}{2} t   \sin (\Deltaz \pwidth t )+\cos (\Deltaz \pwidth t)-1)}{d\;t \sin ^2(\frac{\Deltaz \pwidth}{2} t )+t^3 (\frac{\Deltaz \pwidth}{2})^2}.
\end{align}
In the limit of large dimensions, $d\gg 1$, the rate converges to 
\begin{align}\label{eq:rnd-poisson-gamma-infinite-d}
	\underset{d \rightarrow \infty}{\lim}\gamma_{\rm P}(t) = \frac{2}{t}-\Deltaz \pwidth  \cot \left(\frac{\Deltaz \pwidth}{2} t \right),
\end{align}
which may be compared to the single-qubit decay rate (\ref{eq:single_qubit_uniform_box_decoherence_rate}) for uniformly distributed spectral disorder. In Fig.~\ref{fig:rnd-Poisson}, we show how for different finite dimensions $d$ the depolarization rate decays algebraically while oscillating around the origin with a period $T=(2\pi)/(\Deltaz\pwidth)$. In the high-dimensional limit, we reobtain the periodic divergences already encountered in the case of a single qubit subject to uniform-box-distributed spectral disorder. This underlines that exactly at the times when the rate diverges all coherences completely vanish before undergoing a subsequent revival (see Fig.~\ref{fig:rnd-Poisson}). Note that the general validity condition (\ref{eq:rnd-short-time-validity}) of the short-time master equation evaluates for the PE to $t \ll 1/(\Deltaz\pwidth)$, scaling inversely proportional to the disorder strength.

The dynamical properties of the ensemble average state for unitarily invariant disorder are captured best in terms of its purity,
\begin{align}\label{eq:rnd-state-purity}
	\Tr{\avgdm^2(t)} = a^2(t)\left(\Tr{\dm^2(0)} - \frac{1}{d}\right) + \frac{1}{d}. 
\end{align}
We remind the reader that, by definition, the purity lies in the interval $[1/d, 1]$, where the lower bound corresponds to the maximally mixed state, and the upper to a pure state. In case of the Poissonian ensemble, the mixing probability, Eq.~(\ref{eq:rndmat a(t)}), is $a_{\rm P}(t) = (1+d\: \text{sinc}^2 \left(\frac{\Deltaz \pwidth}{2} t\right))/(1+d)$. As we can see in the insets of Fig.~\ref{fig:rnd-Poisson}, each time the depolarization rate turns negative a revival of purity occurs. In addition, one finds that for finite dimensions the asymptotic state is not the maximally mixed state ($\Tr{\avgdm^2(t)}=1/d$), in spite of the tendency of the depolarization dynamics to drive any initial state towards the maximally mixed state. Formally speaking, this is due to the degenerate terms in $\frnd (t)$ which lead to the constant factor $1/d$ in (\ref{eq:rndmat computef}). Consequently, the larger the dimension, and hence the more "depolarization directions" there are, the smaller the contribution $1/d$ in (\ref{eq:rndmat computef}) gets, and the closer the state is driven towards the maximally mixed state. Being more precise, the asymptotic limit of the purity is given by
\begin{align} \label{eq:rnd-asymptotic-purity}
\underset{{t \to \infty} }{\lim}\Tr{\avgdm^2(t)} = \frac{d+2+\Tr{\dm^2(0)}}{(1+d)^2} \; ; \ d\geq 2.
\end{align}
Hence, the asymptotic limit of the purity (\ref{eq:rnd-asymptotic-purity}) is completely specified by the Hilbert space dimension and the initial state.
This can be understood as a consequence of the Riemann-Lebesgue lemma, stating that, for all functions $f(x)$ which are absolutely integrable on $\mathbb{R}^d$, $\underset{{t \to \infty} }{\lim} \int_{\mathbb{R}^{d}} dx f(x) \exp(i x t) = 0$ holds. Therefore, for any absolutely integrable eigenvalue distribution $p_{\vec{\enslabel}}$, such as the one of the PE (\ref{eq:rnd-poisson-distribution}),  $\underset{{t \to \infty} }{\lim} \avg{\chi}(t) =1/d$, which leads to the asymptotic state (\ref{eq:rnd-asymptotic-purity}).

\begin{figure}
	\center
	\includegraphics[width=0.9\columnwidth]{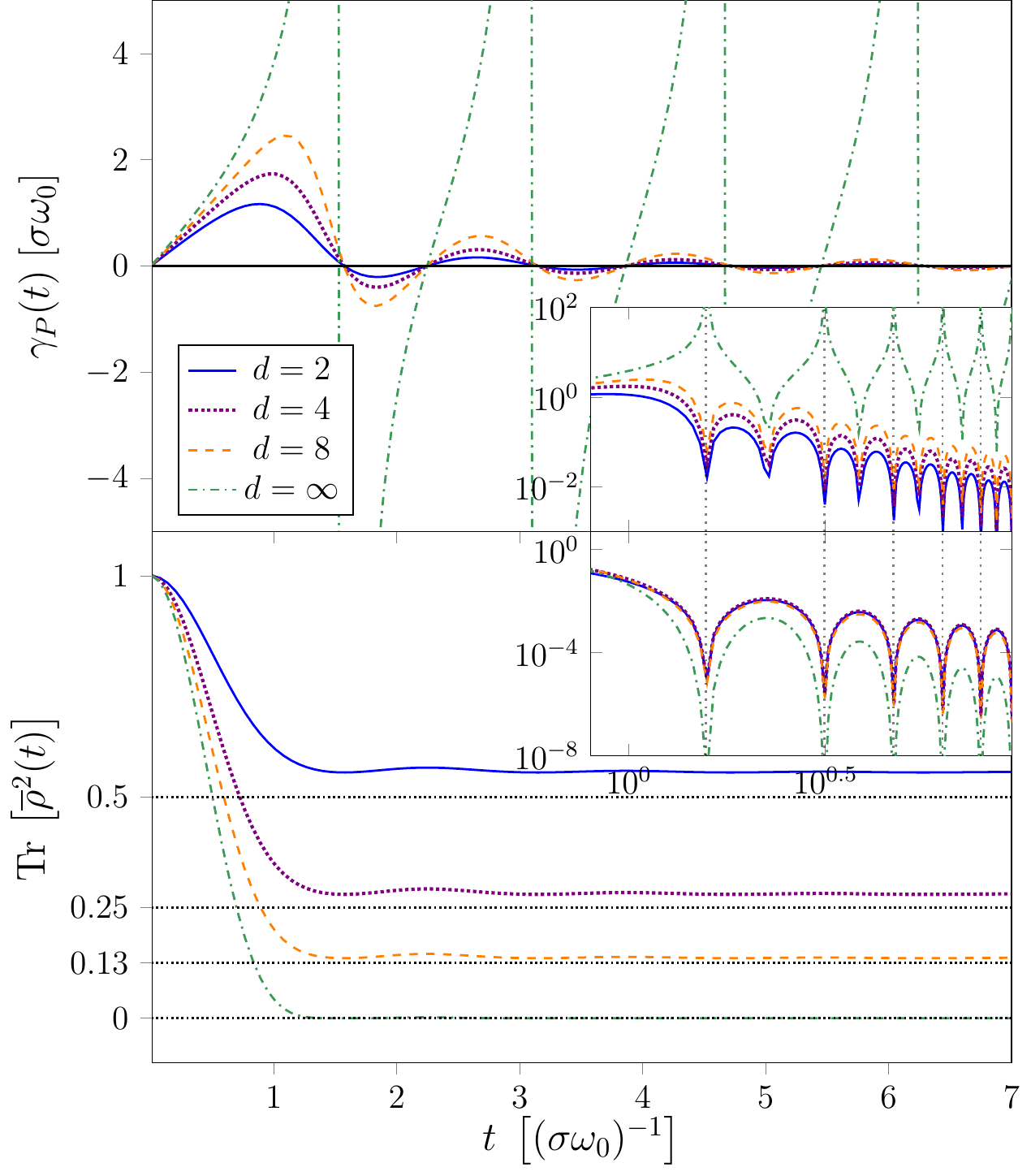}
		\caption{[Color online] Ensemble average dynamics for the unitarily invariant Poissonian disorder ensemble. Top panel: Depolarization rates $\gamma_{\rm P}(t)$ (cf.~Eq.~(\ref{eq:rnd-poisson-rate})) for dimensions $d=2,4,8,\infty$. Inset: Double-logarithmic plot of $|\gamma_{\rm P}(t)|$, showing the ongoing oscillations combined with an algebraic decay to zero for finite dimensions $d=2,4,8$. Bottom panel: Decay of the purity $\Tr{\avgdm^2(t)}$ of a pure initial state with $\Tr{\dm^2 (0)} =1$. For finite dimensions, the purity does asymptotically not drop to the maximally mixed value $1/d$ (indicated by horizontal, black, dotted lines), see also Eq.~(\ref{eq:rnd-asymptotic-purity}). Inset: Double-logarithmic plot of the approach to the asymptotic purity $|\Tr{\avgdm(t)^2}-\underset{t\rightarrow\infty}{\lim}\Tr{\avgdm(t)^2}|$, showing that the purity oscillations are directly correlated with the oscillations of the rate $\gamma_{\rm P}(t)$ around zero. We chose the uniform box distribution width $\pwidth=4$ 
for comparison with the GUE as in \cite{Znidaric2011}.}
		\label{fig:rnd-Poisson}
\end{figure}

\subsection{Gaussian unitary ensemble}

We consider as a second example the Gaussian unitary random matrix ensemble. The joint eigenvalue probability distribution for the GUE \cite{Stockmann2007} reads
\begin{align}\label{eq:rnd-gaussian-distribution}
	p_{\vec{\enslabel}} = \text{const} \cdot \left(e^{-A\sum_{j=1}^d \enslabel_j ^2} \right)\underset{1\leq j < k \leq d}{\Pi}\abs{\enslabel_j-\enslabel_k}^2,
\end{align}
where $A$ is a normalization constant fixing the average density of states at small energies. Following the conventions in \cite{Gessner2013,Znidaric2011}, we set the normalization so that $\avg{\H_{jk}^2}/(\hbar\omega_0)^2=\frac{1}{d}$ (thereby fixing the variance). Since the Gaussian distribution (\ref{eq:rnd-gaussian-distribution}) is invariant with respect to permutations of the eigenvalues $\enslabel_j$, we follow the same procedure as in the previous section in order to compute the sum $\frnd(t)$ of the level-spacing characteristic functions, cf.~(\ref{eq:rndmat computef}). The two-point correlation function for the GUE can be expressed as \cite{Gessner2013,Znidaric2011}
 \begin{align}
 	R_2(\enslabel_1,\enslabel_2)= \text{det} \left[\sum_{n=0}^{d-1} \varphi_n(\enslabel_j)\varphi_n(\enslabel_k)\right]_{j,k=1,2},
 \end{align}
where $\varphi_n(x) = (2^n n! \sqrt{2\pi/d})^{-1/2}e^{-x^2/2} H_n(\sqrt{\frac{d}{2}}x)$ are the harmonic oscillator eigenfunctions, with the Hermite polynomials $H_n(x) \define \exp\left[x^2\right]\left(-d/dx\right)^n \exp\left[-x^2\right]$.

Inserting $R_2(\enslabel_1,\enslabel_2)$ into Eq.~(\ref{eq:rndmat computef}), one can evaluate $\frnd(t)$ for finite dimensions $d$. This then allows one to compute the depolarization rate $\gamma_{\rm GUE}(t)$, Eq.~(\ref{eq:rnd rate}), and the GUE mixing probability $a_{\rm GUE}(t)$, Eq.~(\ref{eq:rndmat a(t)}), specifying the dynamics of the state purity (\ref{eq:rnd-state-purity}). For example, for $d=2$ the depolarization rate reads
\begin{align}
	\gamma_{\rm GUE}(t)|_{d=2}= -\frac{2 \Deltaz^2 t \left(\Deltaz^2t^2-3\right)}{-2 \Deltaz^2t^2+e^{\frac{\Deltaz^2t^2}{2}}+2}.
\end{align}
For the high-dimensional limit, we use the approximation $\frnd(t)|_{d\gg 1} \approx \left( \frac{J_1(2\Deltaz t)}{\Deltaz t}\right)^2$ \cite{Znidaric2011}, where $J_1(t) \define \frac{1}{2\pi} \int_{-\pi}^\pi dx 	\exp \left[i(x-t\sin(x))\right]$ denotes the Bessel function of the first kind, in order to evaluate the GUE depolarization rate $\gamma_{\rm GUE} (t)$ and the mixing probability $a_{\rm GUE}(t)$. 

In Fig.~\ref{fig:rnd-GUE}, we show the time evolution of the GUE depolarization rate $\gamma_{\rm GUE}(t)$ for dimensions $d=2,4,8,\infty$. For finite dimensions, $\gamma_{\rm GUE}(t)$ first exhibits algebraically damped oscillations, before approaching the $x$-axis exponentially from below. In the limit of infinite dimensions, the GUE rate follows a similar temporal evolution as the PE rate, Eq.~(\ref{eq:rnd-poisson-gamma-infinite-d}), including periodic divergences. This reflects a universal property of random matrix ensembles: in the limit of high dimensions, the eigenvalue statistics only depend on the density of states and the type of symmetries fulfilled by the random matrices \cite{Pastur1997} (in the present case, the unitary invariance).

In Fig.~\ref{fig:rnd-GUE} we also show the time evolution of the purity (\ref{eq:rnd-state-purity}). For finite dimensions, the purity initially decays to the minimum $1/d$ of the maximally mixed state, in contrast to the Poissonian case (see Fig.~\ref{fig:rnd-Poisson}). Afterwards, it increases again towards its asymptotic value (\ref{eq:rnd-asymptotic-purity}). The generic purity revivals at short times, which are clearly visible in the insets of Fig.~\ref{fig:rnd-GUE}, are directly related to the oscillations of the depolarization rate $\gamma_{\rm GUE}(t)$ (for more details see Appendix \ref{app:rndmat}). Additionally, we find that the purities approach their asymptotic values exponentially, free of oscillations, from the point when the corresponding decoherence rates decay exponentially to zero. In the high-dimensional limit, we recover a similar behavior as for the Poissonian ensemble, including an infinite number of purity revivals and an algebraic decay towards the asymptotic state. Finally, 
the asymptotic purities of the GUE coincide, because of the absolute integrability of the eigenvalue distribution of the GUE (\ref{eq:rnd-gaussian-distribution}) and due to the Riemann-Lebesgue Lemma, for all dimensions with the values for the PE (c.f. Eq.~(\ref{eq:rnd-asymptotic-purity})).

In summary, systems subject to unitarily invariant disorder all lead to the same asymptotic purity, but following different transient paths characterized by the different spectral disorder distributions. For finite dimensions, the regular Poissonian ensemble leads to an overall algebraic decay towards the maximally mixed state, in contrast to the exponential decay of the chaotic Gaussian ensemble. Similar behavior was noted in the semiclassical description of the decay of autocorrelation functions of chaotic and regular quantum systems \cite{Blumel1988}. 

\begin{figure}
	\center
	\includegraphics[width=0.9\columnwidth]{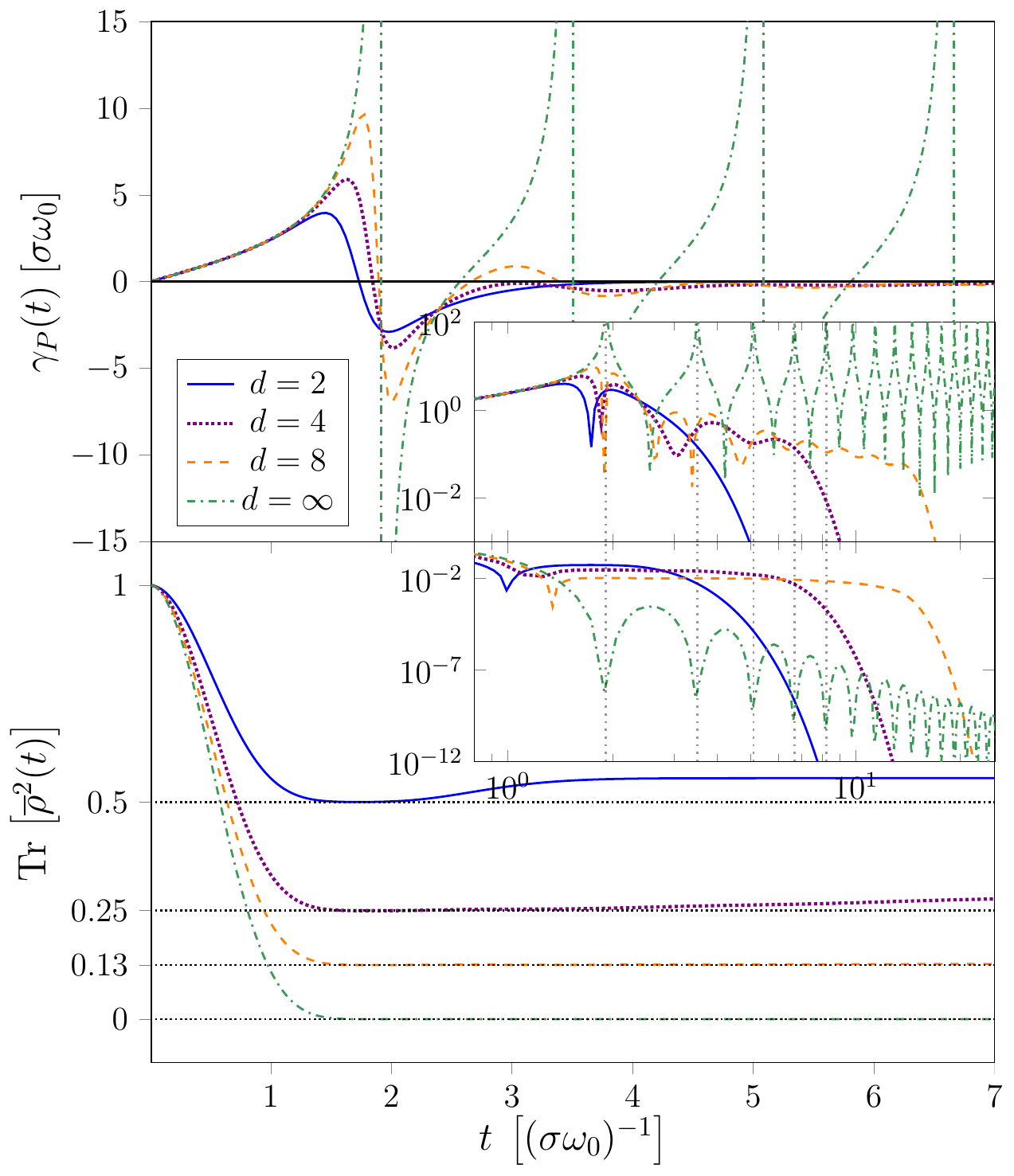}
	\caption{[Color online] Ensemble average dynamics for the unitarily invariant Gaussian disorder ensemble. Top panel: GUE depolarization rate $\gamma_{\rm GUE}(t)$, (cf. \ref{eq:rnd rate}), for dimensions $d=2,4,8,\infty$. Inset: Double-logarithmic plot of $\gamma_{\rm GUE}(t)$. For finite dimensions, the rate undergoes a finite number of oscillations and a subsequent exponential decay to zero from below. Bottom panel:  Decay of the purity $\text{Tr}\left[\avgdm^2(t)\right]$ for a pure initial state, $\text{Tr}\left[\dm^2(0)\right]=1$. For finite dimensions, the asymptotic values of the purities correspond to Eq.~(\ref{eq:rnd-asymptotic-purity}), which is larger than $1/d$ (indicated by horizontal, black, dotted lines). Inset: Double-logarithmic plot of the approach to the asymptotic state $\text{Tr}\left[\avgdm^2(t)\right]-\underset{t\rightarrow\infty}{\lim}\text{Tr}\left[\avgdm^2(t)\right]$. The purity oscillations are directly correlated with the oscillations of the rates $\gamma_{\rm GUE}(t)$ (dotted, vertical, gray lines). Note that, for finite dimensions, the peaks are shifted w.r.t.~their corresponding rates $\gamma_{\rm GUE}(t)$ in the top inset, because of the subtraction of the asymptotic purity value.}
	\label{fig:rnd-GUE}
\end{figure}


\section{Conclusions} \label{sec:conclusions}

We introduced quantum master equations as natural and viable means to conceptually grasp the incoherent effective dynamics of quantum systems described by an ensemble of time-independent, disordered Hamiltonians. To this end, we presented a general method to derive these master equations as they emerge from the disorder average. Therewith, we provided a new tool to study the average dynamics of disordered quantum systems both on transient and asymptotic time scales. Our approach furthermore naturally separates coherent and incoherent components of the dynamics, and associates characteristic time and coupling constants of the effective time evolution with the statistical features of the generating disorder.

In the paradigmatic case of spectral disorder, we found that the common basis of eigenstates of the ensemble Hamiltonians specifies the decoherence process, i.e.,~dephasing, while the eigenvalue distributions determine the corresponding time-dependent decoherence rates. Similarly, in the illustrative case of unitarily invariant disorder, we obtained that the isotropic eigenstate distribution leads to a depolarization process with a rate again given by the eigenvalue distribution. These two cases were solved exactly. It is clear that treating more complex and realistic disorder models will require some degree of approximation. For example, the master-equation in Lindblad form presented in Section \ref{sec:short-time_master_equation}, which is valid in the limit of short times, has been used in \cite{Gneiting2016} to study homogeneous disorder models.
In order to extend its validity beyond short times, one could employ a Born-Markov-type approximation, as done in \cite{Mueller2009} to derive an evolution equation for diffusive spin-transport in a disordered medium. Another perspective to describe complex disordered system may consist in complementing our master equation approach with techniques from disorder transport theory such as perturbation theory or diagrammatic methods as used, e.g., in the study of weak localization \cite{Wellens2008a}. Nevertheless, the generic examples here treated in detail clearly illustrate that the master equation description of the ensemble-averaged dynamics is not only a new approach to the dynamics of disordered systems, but also provides useful and non-trivial information, e.g., on the coherent and incoherent contributions to the effective dynamics, respectively. The latter cannot be readily extracted from the asymptotic, disorder-averaged state alone.

From a more general point of view, we conclude that the character of the decoherence process is determined by the structure of the disorder, i.e.,~the eigenstate distribution, while the temporal course of the master equation follows from the eigenvalue distribution. The latter then mediate dynamical features such as revivals. Therefore, knowing these master equations, we are able to characterize concisely the collective dynamics of the ensemble average. Conversely, their experimental inference may allow one to identify the underlying disorder properties. In other words, the experimental monitoring of the effective dynamics on transient time scales could be used as a diagnostic tool to uncover the characteristic features of the disorder with the help of the master equations introduced in this article.

The necessity to average over an ensemble of disordered Hamiltonians constitutes a ubiquitous issue in the quantum theory of 
complex systems, be it for experimental or for conceptual reasons, and to this end the presented master equation description may be decisive in order to obtain a thorough understanding of the associated dynamics. Besides the discussed occurrence of spectral disorder in trapped ion systems exposed to fluctuating external magnetic fields \cite{Gessner2014}, we would like to mention, as other interesting experimental examples, the transport of electrons or excitations in biomolecular systems subject to slow conformational changes \cite{Kruger2011,Walschaers2015}, cold atom implementations of quantum walks, where residual thermal fluctuations lead to an effective differential light shift \cite{Alberti2014a}, and the propagation of photons in the presence of a turbulent atmosphere, giving rise to decoherence in the orbital angular momentum degree of freedom \cite{Roux2014b,Roux2015}. On the conceptual side, we mention the characterization of the ensemble average dynamics in the seminal Anderson model \cite{Anderson1958}, coherent backscattering \cite{Corey1995,Barabanenkov1973}, and the identification of robust features in boson sampling systems \citep{Aaronson2013} based on random matrices \cite{Walschaers2014}.

Besides understanding the dynamical impact of averaging over an ensemble, the master equation approach also suggests that disorder can serve as a useful resource to generate quantum dynamical maps of interest. Each disorder master equation gives rise to a family of random unitary maps, which are important tools in quantum information theory \cite{Chruscinski2014a,Hall2014,Breuer2012,Nielsen2000,Bengtsson2006}. More generally speaking, it may be insightful to clarify the relationship between the ensemble average dynamics of disordered systems and the reduced dynamics of open systems, i.e.,~to identify environmental models giving rise to the same system dynamics as the ensemble average. This will help to clarify whether encountered dynamical properties must be traced back to disorder or may also be a consequence of the coupling to an environment. Our approach thus establishes a non-trivial connection between the theory of disordered and open quantum systems, and of quantum information.

\section*{Acknowledgement} C.M.K. acknowledges funding by the German National Academic Foundation. A.B. acknowledges financial support from the EU Collaborative project QuProCS (Grant Agreement 641277). The authors thank Cord A. Müller for helpful comments on the manuscript The article processing charge was funded by the German Research Foundation (DFG) and the University of Freiburg in the funding programme Open Access Publishing.


\appendix

\section{Characteristic function and cumulant expansion}\label{app: characteristic fct}
The characteristic function $\phi(t)$ \cite{Lukacs1972} of a probability distribution $p(x)$ is defined as
\begin{align}
	\phi(t) &\define \int_{-\infty}^\infty dx\, p(x) e^{ixt}. 
\end{align}
The cumulant expansion of the probability density $p(x)$ \cite{Kenney1947} is then defined as
\begin{align}
		\ln\left[\phi_x (t) \right] &\define \sum_{n=1}^\infty \kappa^{(n)} \frac{(it)^n}{n!},
\end{align}
with $\kappa^{(n)}$ the $n$th cumulant.

The cumulants are related to the raw moments of $p(x)$, $\mu^{(n)} = \int_{-\infty}^\infty dx \, x^n p(x) $,  via
\begin{align}
	\kappa^{(n)} = \mu^{(n)} - \sum_{m=1}^{n-1}\binom{n-1}{m-1} \kappa^{(m)}\mu^{(n-m)}.
\end{align}
For the first few cumulants one obtains
\begin{align}
\kappa^{(1)} &=\mu^{(1)} \\
\kappa^{(2)} &=\mu^{(2)} -{\mu^{(1)}}^2 \\
\kappa^{(3)} &=\mu^{(3)}-3\mu^{(2)}\mu^{(1)} +2{\mu^{(1)}}^3\\
\kappa^{(4)} &=\mu^{(4)}-4\mu^{(3)}\mu^{(1)} -3{\mu^{(2)}}^2+12\mu^{(2)}{\mu^{(1)}}^2-6\mu^{(1)}.
\end{align}
The first raw moment $\mu^{(1)} $ is usually referred to as the mean and $\mu^{(2)} - {\mu^{(1)}}^2$ as the variance.

\section{Coherent dynamics of a single qubit with spectral disorder}\label{app:qubit-coherent}

In Figure \ref{fig:qubit-all}, the decay of the norm of the coherences of the ensemble average state in the case of a single qubit subject to spectral disorder was shown for the four, Cauchy-Lorentz, Gaussian, uniform box, and L\'evy distributions. This highlighted the effect of the incoherent part of the master equation (\ref{eq:qubitqme}), namely the decoherence rate, upon the dynamics. In Figure \ref{fig:qubit-all-coherent}, we show the time evolution of the coherences, Eq.~(\ref{eq:qubit-dynamics-solved}), in the Bloch sphere representation, which allows us to better appreciate the combined effects of the coherent and incoherent part of the master equation upon the dynamics, which were extensively discussed in Section \ref{sec:diagdisorder}. Summarizing the previous discussions, we have that the Cauchy-Lorentz and the Gaussian distributions lead to exponential and Gaussian dephasing, respectively. The dynamics arising from the uniform box distribution show a slower, algebraic dephasing, and in addition lead to periodic passages through the maximally mixed state and subsequent revivals. Finally, the L\'evy distribution gives also rise to an algebraic dephasing and moreover leads to an accelerating coherent precession frequency as compared to the constant precession frequency for the three other distributions.

\begin{figure}
		\includegraphics[width=0.9\columnwidth]{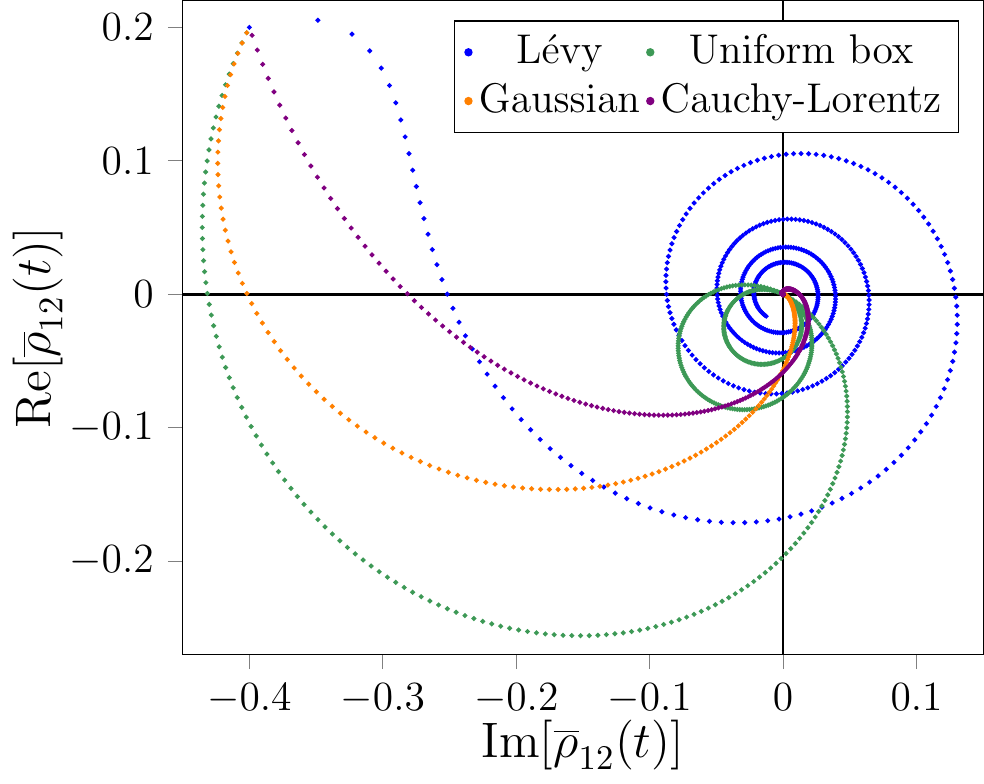}
			 \caption{[Color online] Ensemble average dynamics of a single qubit with spectral disorder. Shown are the dynamics of the coherences in the Bloch sphere representation, i.e.,~the real and imaginary part of the off-diagonal terms of the average density matrix $\avgdm_{12}(t)=\avgdm^*_{21}(t)$ (c.f. Eq.(\ref{eq:qubit-dynamics-solved})), for the Cauchy-Lorentz, Gaussian and L\'evy distributions with coinciding scale parameter $\pwidth$, and, for visual clarity, for the uniform box distribution with $2\pwidth$. The initial parameters are chosen as in Figure \ref{fig:qubit-all}, yielding $\dm_{12}(0)=1/5-i\,2/5$. Time runs from $t\in\left[0,3\pi\right]$ in units of $(\pwidth \Deltaz)^{-1}$. All four distributions lead to dephasing dynamics, as reflected by the Lindblad operators of the disorder master equation (\ref{eq:qubitqme}), but follow distinct trajectories determined by the respective decoherence rates and energy functions (\ref{eq:qubit-lorentz-rates}),(\ref{eq:qubit-rates-gaussian}),(\ref{eq:single_qubit_uniform_box_decoherence_rate}) and (\ref{eq:qubit-levy-rates}). In particular, one can see the contribution of the time-dependent part of the energy function for the L\'evy distribution (\ref{eq:qubit-levy-rates}) in the accelerating coherent precession frequency (blue line), as compared to the constant precession frequency for the three other distributions.}
	 \label{fig:qubit-all-coherent}
\end{figure}

\section{GUE short-time dynamics of the purity}\label{app:rndmat}

In Fig.~\ref{fig:rnd-GUE}, it was shown that the ensemble average purity for the Gaussian unitarily invariant disorder ensemble first decays to the maximally mixed value of $1/d$ before reaching the asymptotic state purity (\ref{eq:rnd-asymptotic-purity}) from below. In Fig.~\ref{fig:purityguelog}, we illustrate in more detail how the dynamics of the purity follows from the depolarization rate $|\gamma_{\rm GUE}(t)|$ at short times. The oscillations of $\gamma_{\rm GUE}(t)$ can clearly be related to the oscillations of the purity. For all dimensions, the purity first decays to the maximally mixed value $1/d$. For large times and finite dimensions, the rate is actually negative (see Fig.~\ref{fig:rnd-GUE}), and thus the purity eventually increases to the asymptotic value (\ref{eq:rnd-asymptotic-purity}). In the high-dimensional limit, the rate remains periodic for all times, inducing an infinite number of purity revivals, with an algebraically decaying envelope.
\begin{figure}[H]
		\includegraphics[width=0.9\columnwidth]{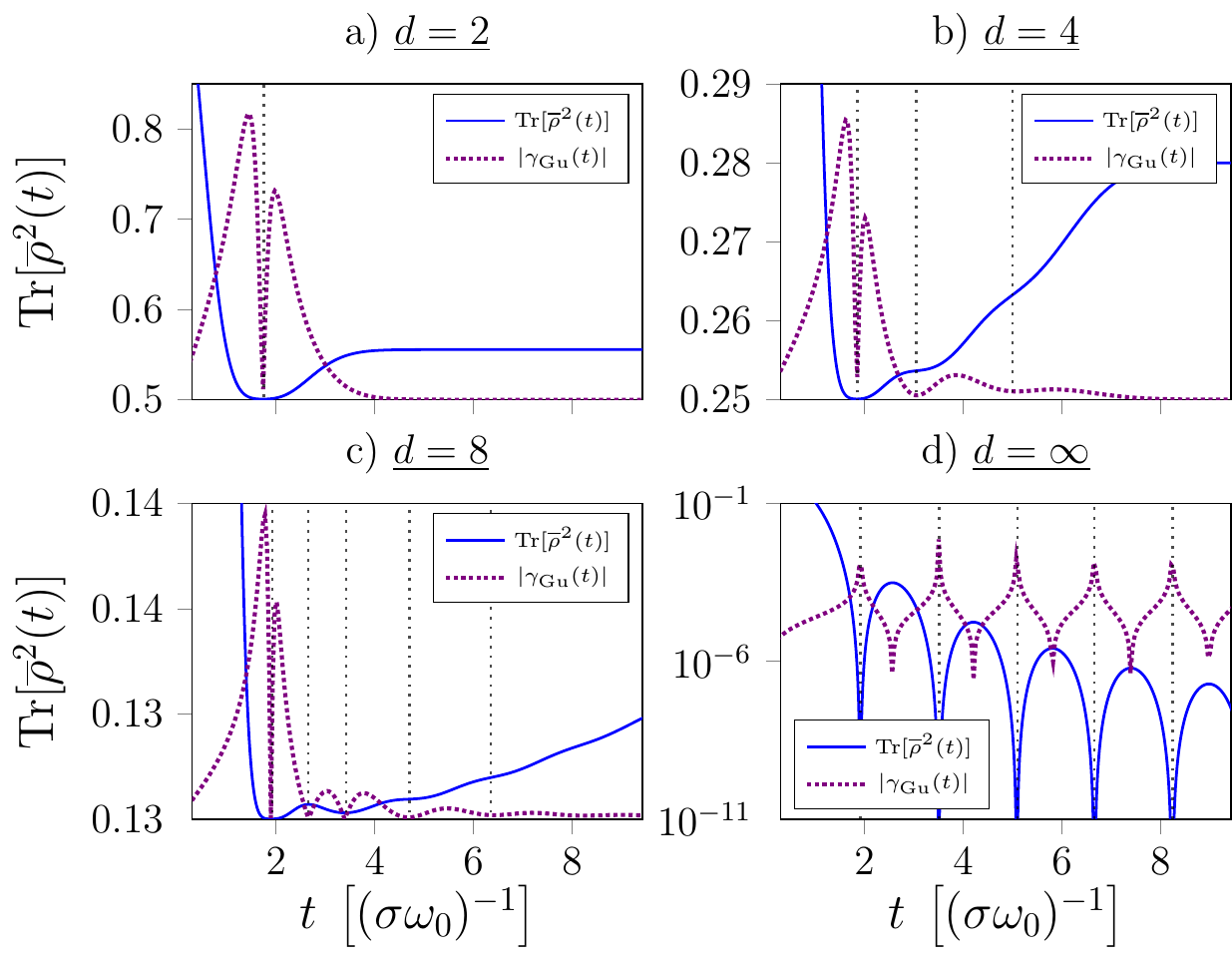}
			\caption{\label{fig:purityguelog}[Color online] Decay of the purity $\text{Tr}\left[\avgdm^2(t)\right]$ to its asymptotic value (blue solid) and of the absolute value of the depolarization rates $|\gamma_{\rm GUE}(t)|$ (purple dashed), for the GUE ensemble for dimensions a) $d=2$, b) $d=4$ and c) $d=8$. For dimension d) $d\rightarrow\infty$, the plot is semi-logarithmic. All the depolarization rates $|\gamma_{\rm GUE}(t)|$ are rescaled and vertically shifted for visualization purposes. The vertical, gray, dotted lines are indicating the purity revivals.}
\end{figure}

\section{Equivalence of time-local and non-local form}\label{app:qme equivalence}

We present the essential steps showing that master equations in local form (c.f. Eq. (\ref{eq:qme})) and master equations with a memory kernel are equivalent to one another \cite{Chruscinski2010, Andersson2007}. 

Suppose the dynamical map $\dm(t) = \tmap_{t-t_0}\left[\dm(t_0)\right]$ is generated by an integro-differential master equation with memory kernel $\mathcal{K}_s(t-t_0)$
\begin{align}
	\frac{d}{dt}\dm (t-t_0)= \int_{t_0}^t ds\, \mathcal{K}_s(t-t_0) \left[\rho(s)\right].
\end{align}
Using the formal inverse of the dynamical map, $\tmap_{t-t_0}^{-1}$, we obtain
\begin{align}
	\frac{d}{dt}\dm (t-t_0)&= \int_{t_0}^t ds\, \mathcal{K}_s(t-t_0) \left[\rho(s)\right] \\
	&= \int_{t_0}^t ds\, \mathcal{K}_s(t-t_0) \circ \tmap_s  \left[\rho(t_0)\right]  \\
	&= \int_{t_0}^t ds\, \mathcal{K}_s(t-t_0) \circ \tmap_s \circ \tmap_{t-t_0}^{-1} \left[\tmap_{t-t_0}\rho(t_0)\right] \\
	&=\dot{\tmap}_{t-t_0} \tmap_{t-t_0}^{-1} \left[\dm(t-t_0)\right] \label{eq:time-local form},
\end{align}
with the map $\dot{\tmap}_{t-t_0} = \int_{t_0}^t ds\, \mathcal{K}_s(t-t_0) \circ \tmap_s$ being a function of $t-t_0$ only (and not of $s$ anymore). The last equality (\ref{eq:time-local form}) is precisely our time-local master equation (\ref{eq:formal qme}). Therefore, the intergro-differential form and the time-local form are equivalent, and the latter is perfectly able to describe memory effects. This is essentially because the time-local generator explicitly depends on the time-difference $t-t_0$ and not only on $t$ (this can easily be overlooked as one typically sets $t_0=0$). On a technical level, going from one form to another can for example be done by taking the Laplace transform.
%
%
%
%


\bibliographystyle{apsrev4-1}
\bibliography{bibliography,footnotes}

\end{document}